\documentclass[twocolumn, graphics, floatfix, a4paper, aps, prx,
superscriptaddress, longbibliography, showpacs,citeautoscript]{revtex4-2}

\usepackage{graphicx}
\usepackage{amsmath}
\usepackage{amssymb}
\usepackage{color}
\usepackage{ulem}
\usepackage[dvipsnames]{xcolor}
\usepackage[breaklinks=true,colorlinks=true,urlcolor=blue,citecolor=blue,linkcolor = blue]{hyperref}

\renewcommand{\vec}[1]{\boldsymbol{#1}}
\newcommand{\up}{\uparrow}
\newcommand{\dn}{\downarrow}
\newcommand{\crea}[1]{#1^{\dag}}
\newcommand{\ani}[1]{#1^{\vphantom{\dag}}}
\newcommand{\ave}[1]{\left\langle #1\right\rangle}
\newcommand{\bra}[1]{\langle #1|}
\newcommand{\ket}[1]{|#1\rangle}

\newcommand{\bea}{\begin{equation} \begin{aligned}}
\newcommand{\eea}{\end{aligned} \end{equation} }
\newcommand{\be}{\beta}
\newcommand{\al}{\alpha}
\DeclareRobustCommand{\Eq}[1]{Eq.~(\ref{#1})}
\newcommand{\eps}{\epsilon}

\renewcommand{\u}{\uparrow}
\renewcommand{\d}{\downarrow}
\newcommand{\del}{\partial}

\newcommand{\kk}{{\vec{k}}}
\newcommand{\RR}{{\vec{R}}}
\newcommand{\tL}{{\tilde L}}
\newcommand{\KK}{{\vec{K}}}

\begin{document}

\title{Revisiting flat band superconductivity: dependence on minimal
  quantum metric and band touchings}
\author{Kukka-Emilia Huhtinen}
\email{kukka-emilia.huhtinen@aalto.fi}
\affiliation{Department of Applied Physics, Aalto University School of Science,
FI-00076 Aalto, Finland}
\author{Jonah Herzog-Arbeitman}
\affiliation{Department of Physics, Princeton University, USA}
\author{Aaron Chew}
\affiliation{Department of Physics, Princeton University, USA}
\author{Bogdan A. Bernevig}
\affiliation{Department of Physics, Princeton University, USA}
\affiliation{Donostia International Physics Center, P. Manuel de Lardizabal 4, 20018 Donostia-San Sebastian, Spain}
\affiliation{IKERBASQUE, Basque Foundation for Science, Bilbao, Spain}
\author{P\"aivi T\"orm\"a}
\email{paivi.torma@aalto.fi}
\affiliation{Department of Applied Physics, Aalto University School of Science,
FI-00076 Aalto, Finland}

\date{\today}

\begin{abstract}
  
   A critical result in superconductivity is that flat bands, though dispersionless, can still host nonzero superfluid weight due to quantum geometry. We show that the derivation of the mean field superfluid weight in previous literature is incomplete, which can lead to severe quantitative and even qualitative errors. We derive the complete equations and demonstrate that the minimal quantum metric---  the metric with minimum trace --- is related to the superfluid weight in isolated flat bands. We complement this result with an exact calculation of the Cooper pair mass in attractive Hubbard models with the uniform pairing condition. When the orbitals are located at high symmetry positions, the Cooper pair mass is exactly given by the quantum metric, which is guaranteed to be minimal. Moreover, we study the effect of closing the band gap between the flat and dispersive bands, and develop a mean-field theory of pairing for different band-touching points via the $S$-matrix construction. In mean field, we show that a non-isolated flat band can actually be beneficial for superconductivity. This is a promising result in the search for high temperature  superconductivity as the material does not need to have flat bands that are isolated from other bands by the thermal energy. Our work resolves a fundamental caveat in understanding the relation of multiband superconductivity to quantum geometry, and the results on band touchings widen the class of systems advantageous for the search of high temperature flat band superconductivity. 

\end{abstract}

\maketitle

\section{Introduction}

\begin{figure}
  \centering
  \includegraphics[width=\columnwidth]{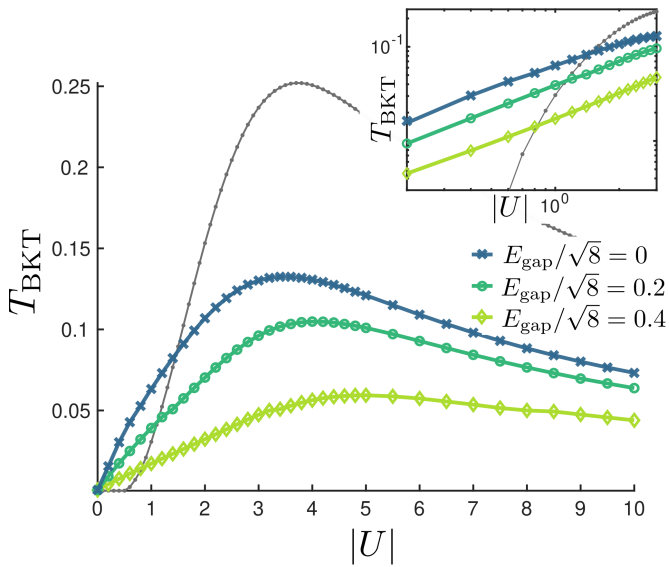}
  \caption{BKT temperature computed for the square lattice (gray) and for the Lieb lattice with a
    half-filled flat band (blue, green and yellow) with different values of the
    hopping staggering (see Fig.~\ref{fig.example}). Inset: BKT temperature at interactions $0.2\leq |U|\leq 3$. The flat band is
    isolated from the other bands by a band gap $E_{\rm gap} =
    \sqrt{8}\delta$. The highest BKT temperatures are obtained when
    $\delta=0$, corresponding to the  situation where the gap between
    the flat band and dispersive bands closes, resulting in a
    linear band touching. The BKT temperature for the square lattice
    (gray) is exponentially suppressed at low interactions, whereas
    $T_{BKT}$ on the isolated flat band is proportional to $|U|$. All energies are given in units of the average inter-lattice-site hopping energy $t$.} \label{fig.comp} 
\end{figure}

Systems with dispersionless (flat) bands host exotic phenomena, as even small interactions will dominate the kinetic energy. For example, flat bands have been predicted to increase the critical temperature for superconductivity. Bardeen-Cooper-Schrieffer (BCS) theory predicts that the critical temperature is given by $T_c\propto {\rm exp}\left(-\frac{1}{|U| \rho_0(E_F)}\right)$, where $|U|$ is the strength of the effective attractive interaction and $\rho_0(E_F)$ is the density of states at the Fermi surface. In a flat band, where the density of states diverges, $T_c$ is proportional~\cite{Heikkila2011,Khodel1990,Kopnin2011} to $|U|$, implying that the critical temperature can be much higher in flat bands than in dispersive bands at low interaction strengths. 

However, the BCS critical temperature does not by itself indicate superconductivity, as it is only the critical temperature for Cooper pair formation. The Meissner effect and the possibility of dissipationless transport are also required. These are characterized by a nonzero superfluid weight $D_s$ or, equivalently, superfluid stiffness~\cite{Scalapino1993}. Moreover, a nonzero superfluid weight is a necessary condition for a nonzero Berezinsky-Kosterlitz-Thouless (BKT) transition temperature, which is the critical temperature for superconductivity in two dimensions. The superfluid weight is conventionally given by $D_s=n_e/m^*$, where $n_e$ is the total particle density and $m^*$ is the effective mass. In a flat band, single particles localize and $m^*$ diverges, which indicates vanishing superfluid weight. However, in multiband models, the superfluid weight has an additional geometric contribution which can be nonzero even in the case of flat bands~\cite{Peotta2015,Liang2017,Julku2016}. In the isolated band limit, this contribution has been shown~\cite{Peotta2015} to be related to the quantum metric~\cite{Provost:1980,Resta2011,OzawaGoldman2018}. Monte Carlo results are in good agreement with this prediction~\cite{Hofmann2020,Herzog-Arbeitman2021,Peri2021}. Flat band superconductivity has attracted immense interest due to its relevance in magic-angle twisted bilayer graphene~\cite{Cao2018,Cao2018b,Yankowitz2019} and other moiré materials~\cite{Park2021,Shen2020,Cao2020,Balents2020,Andrei2021}. In particular, the potential importance of the geometric contribution to the superfluid weight has been shown in theoretical studies of twisted bilayer graphene~\cite{Xie2020,Julku2020,Hu2019,Torma2021}, and has also been explored experimentally~\cite{Tian2021}.

There is, however, a fundamental problem in the relation between the superfluid
weight and the quantum metric as presented in previous literature. Consider a gedanken transformation that changes the orbital locations of a lattice model without altering the hopping terms. 
The superfluid weight is invariant under such
transformations. On the other hand, the quantum metric depends not
only on the tight-binding parameters of the lattice model, but also on
the locations of the orbitals. We show that this discrepancy in
mean-field theory is resolved by properly accounting for the
dependence of the order parameters on the magnetic vector potential. This dependence is
crucial in multiband models, where the order parameters in different
orbitals can have different complex phases. We show that
accounting for the behavior of the order parameters is necessary even
in systems with time-reversal symmetry and uniform pairing, contradicting previous
literature~\cite{Peotta2015,Chan2022}. We derive complete equations
for the mean-field superfluid weight, and show that the use of the
simpler equations provided in previous literature can lead to
quantitative and, in extreme cases, qualitative errors where
the superfluid weight is incorrectly nonzero. Within our general mean-field framework, we study lattice models with both isolated and
non-isolated flat bands. We show that, in time-reversal symmetric systems, the superfluid weight for
isolated flat bands is proportional to the \textit{minimal} quantum metric,
which is the quantum metric with the smallest possible trace for the
considered lattice model. 

These conclusions in mean field theory are mirrored by exact calculations of the Cooper pair mass in attractive Hubbard models possessing a uniform pairing condition. We find two contributions to the effective mass in perturbation theory: the quantum metric and a competing non-universal term. However, we show that the space group symmetries strongly constrain the latter. If the orbitals are located at high symmetry positions such that they are pinned in location by the lattice symmetries, then this non-universal term vanishes and the quantum metric is the unique contribution to the Cooper pair mass. We propose a simple extension of the uniform pairing condition that guarantees the non-universal term vanishes.

Based on our results, we conclude that lower bounds for
the superfluid weight in terms of topological invariants such as the
Chern number~\cite{Peotta2015} and Euler class~\cite{Xie2020} are
valid, but the use of other bounds which depend on orbital
positions~\cite{Liang2017,Herzog-Arbeitman2021} requires additional
conditions, e.g.~space group symmetries. In obstructed atomic limits~\cite{Herzog-Arbeitman2021}, the superfluid weight is only bounded by real space invariants computed at the high-symmetry positions. Moreover, we discuss which results in previous literature are
likely to be accurate, and which would need revisiting based on
the complete formula for the superfluid weight.

In order to understand the behavior of non-isolated flat bands, we
also study the effect of closing the gap between the flat band and
dispersive bands. Remarkably, we show that a band touching can actually be beneficial for
superconductivity (see Fig.~\ref{fig.comp}). This is important, as it means that one does not need to find systems where the flat band is separated from the other bands by a large energy scale. If isolated bands were needed, trying to achieve a higher critical temperature would mean that larger band gaps were required to avoid thermal excitations to the other bands --- this could be a severe limitation especially when searching for room temperature superconductivity. Our results show that such isolation is not necessarily needed. In contrast, band touchings can enhance $T_{BKT}$ or $T_c$. 

We also investigate the
effect of different types of band touchings, and show that the quantum
geometry of the flat band alone is not sufficient to describe
superconductivity in the non-isolated band limit: the type of band touching matters too, and can actually be used as a design degree of freedom when optimizing the critical temperature. We complement our numerical results with an analytic treatment of interacting bipartite crystalline lattices with mean field theory, yielding relations between the pairing strengths on different sublattices.  

Overall, our results are promising for harnessing the potential of flat bands in increasing the critical temperature of superconductivity. This potential is illustrated by Fig.~\ref{fig.comp}. For large interactions, dispersive band structures are often as good or better than flat band systems. In contrast, for weak interactions (typically $|U| < t$), flat bands provide a clear, even radical, advantage. This makes it possible to utilize a wider class of systems and materials for high temperature superconductivity since interactions do not need to be strong. The potential of flat bands to offer high critical temperature even for weak interactions may also help avoid bipolarons and charge density waves competing with superconductivity at large interactions~\cite{Esterlis2018,Esterlis20182}.  

This article is structured as follows. In Sec~\ref{sec.sfw}, we derive
the complete equations for the superfluid weight and show how they
differ from the results obtained in previous literature. We then
revisit superconductivity in isolated flat bands in Sec.~\ref{sec.qm},
and show that the superfluid weight is related to the minimal quantum
metric. Sec.~\ref{sec.example} illustrates our general results within the specific example of the Lieb lattice. In Sec.~\ref{sec.Jonah}, we show how the general conclusions given by the superfluid weight calculations can be obtained by derivation of the many-body effective Cooper pair mass in a flat band, and how symmetries can guarantee that the quantum metric is minimal. In Sec.~\ref{sec.ni}, we study non-isolated flat bands, and
show that the highest $T_{BKT}$ can occur when the flat band is \textit{not}
isolated from the dispersive bands. The validity of results given in previous literature is discussed in Sec.~\ref{sec.prev}. Finally, we summarize our conclusions in Sec.~\ref{sec.conc}.

\section{Superfluid weight in multiband mean-field
  models} \label{sec.sfw} 

\subsection{The model Hamiltonian}

We study the Hubbard model on a multiband lattice
\begin{align}
  H &= \sum_{\sigma}\sum_{i\alpha,j\beta} (t_{i\alpha,j\beta}^{\sigma} -
  \mu \delta_{i\alpha,j\beta})\crea{c_{i\alpha\sigma}}\ani{c_{j\beta\sigma}}
  \nonumber \\
  &+ U\sum_{i\alpha} \crea{c_{i\alpha\up}} \crea{c_{i\alpha\dn}}
  \ani{c_{i\alpha\dn}} \ani{c_{i\alpha\up}}, \label{eq.ham}
\end{align}
where $i,j$ label the unit cells and $\alpha,\beta$ the orbitals in a
unit cell. The hopping amplitude from site $j\beta$ to $i\alpha$ for
spin $\sigma$ is $t_{i\alpha,j\beta}^\sigma$ and $U<0$ is the on-site
interaction strength. The particle number is tuned by the
chemical potential $\mu$. We use the usual mean-field approximation
$U\crea{c_{i\alpha\up}}\crea{c_{i\alpha\dn}}\ani{c_{i\alpha\dn}}\ani{c_{i\alpha\up}}\approx
\Delta_{i\alpha}\crea{c_{i\alpha\up}}\crea{c_{i\alpha\dn}} + {\rm
  H.c.} - |\Delta_{i\alpha}|^2/U$, where $\Delta_{i\alpha} =
U\ave{c_{i\alpha\dn} c_{i\alpha\up}}$. We will
focus on solutions where the order parameter
is uniform on each orbital, $\Delta_{i\alpha}=\Delta_{\alpha}$, i.e. it does not depend on the unit cell index $i$ but
can depend on the orbital index $\alpha$.

\subsection{Superfluid weight from the free energy} \label{sec.fromfreeenergy}

The superfluid weight can be defined as the change in free energy $F=\Omega + \mu N$, where $\Omega$ is the grand canonical
potential and $N$ is the particle number, due to a change in the phase of
the order parameters $\Delta_{i\alpha}\to \Delta_{i\alpha}
e^{2i\vec{q}\cdot \vec{r}_{i\alpha}}$~\cite{Taylor2006,Peotta2015}, with $\vec{r_{i\alpha}}$ being the position of the site $i\alpha$:
\begin{equation}
  [D_s]_{ij} = \frac{1}{V}\frac{{\rm d}^2F}{{\rm d}q_i{\rm
      d}q_j}\bigg|_{\vec{q}=\vec{0}}. \label{eq.sfw} 
\end{equation}
Here, $V$ is the volume of the system.
The derivative is taken at a constant temperature, but the other
thermodynamic variables are allowed to vary with $\vec{q}$. 

Introducing the phase $e^{2i\vec{q}\cdot \vec{r}_{i\alpha}}$ into Eq.~\eqref{eq.ham}, the Fourier transformed mean-field Hamiltonian reads
\begin{align}
  H(\vec{q}) &= \sum_{\vec{k}} \vec{\crea{c_{\vec{k}}}}
  H_{\rm BdG}(\vec{k}) \vec{\ani{c_{\vec{k}}}} \nonumber \\
  &+ \sum_{\vec{k}}{\rm
    Tr} H_{\vec{k}}^{\dn} - nN_c\mu - N_c\sum_{\alpha}
  \frac{|\Delta_{\alpha}(\vec{q})|^2}{U}, \\
  H_{\rm BdG}(\vec{k}) &= \begin{pmatrix}
    H_{\vec{q}+\vec{k}}^{\up} - \mu \vec{1} & \vec{\Delta} \\
    \vec{\Delta}^{\dag} & - (H_{\vec{q}-\vec{k}}^{\dn})^* + \mu \vec{1}
  \end{pmatrix},
\end{align}
where $\vec{\ani{c_{\vec{k}}}} = (
\ani{c_{\vec{q}+\vec{k},\alpha=1,\up}},\ldots,
\ani{c_{\vec{q}+\vec{k},\alpha=n,\up}}$, $ 
\crea{c_{\vec{q}-\vec{k},\alpha=1,\dn}},\ldots,\crea{c_{\vec{q}-\vec{k},\alpha=n,\dn}})^{\rm  
  T}$ and $n$ is the number of bands. 
  The
number of unit cells is denoted by $N_c$, and 
$\vec{\Delta}={\rm diag}(\Delta_{1},\ldots,\Delta_n)$.
The matrix $H_{\vec{k}}^{\sigma}$ is the Fourier transformation of the
kinetic Hamiltonian for spin $\sigma$,
$[H_{\vec{k}}^{\sigma}]_{\alpha\beta} = \sum_{i}
t_{i\alpha,0\beta}^{\sigma} e^{-i\vec{k}\cdot
  (\vec{R_i}+\vec{\delta_{\alpha}}-\vec{\delta_{\beta}})}$, where
$\vec{R_i}$ is the position of the $i$th unit cell and
$\vec{\delta_{\alpha}} = \vec{r_{i\alpha}}-\vec{R_i}$. 
 Here we have used the Fourier transformation
\begin{equation}
c_{\vec{k}\alpha\sigma} =
\frac{1}{\sqrt{N_c}} \sum_{i}
e^{-i\vec{k}\cdot(\vec{R_i}+\vec{\delta_{\alpha}})} c_{i\alpha\sigma}, \label{eq.fourier}
\end{equation}
which takes the intra-unit cell positions of the orbitals into
account. Another convention that is often used is
\begin{equation}
  c_{\vec{k}\alpha\sigma} =
  \frac{1}{\sqrt{N_c}} \sum_{i}
  e^{-i\vec{k}\cdot\vec{R_i}} c_{i\alpha\sigma}, \label{eq.fourier_no}
\end{equation}
which corresponds to setting all $\vec{\delta_{\alpha}}=\vec{0}$. This
latter convention has the advantage of making the Hamiltonian explicitly periodic
in reciprocal space. However, the choice of the orbital positions
plays an essential role, as we will show, in relating the superfluid weight to quantum geometry.

The equilibrium
state minimizes the grand canonical potential
\begin{align}
  \Omega &= -\frac{1}{\beta} \sum_{\vec{k}}\sum_{i} \ln
         [1+\exp(-\beta E_{\vec{k},i})]  \nonumber \\
         &+\sum_{\vec{k}}{\rm
           Tr}H_{\vec{k}}^{\dn} - nN_{c}\mu - N_c\sum_{\alpha}
         \frac{|\Delta_{\alpha}|^2}{U} , 
\end{align}
where $E_{\vec{k},i}$ are the eigenvalues of the Bogoliubov-de-Gennes
Hamiltonian $H_{\rm BdG}({\vec k})$. The order parameters for a given chemical
potential and temperature can thus be solved by minimizing $\Omega$,
or equivalently by solving the gap equation. The particle number is
controlled by the chemical potential $\mu$, and fulfills the equation
$N = -\partial \Omega/\partial \mu$.

Equation~\eqref{eq.sfw} can be cumbersome to use, as it requires knowledge of the state at nonzero $\vec{q}$. In previous
literature~\cite{Peotta2015}, it has been shown that this equation
simplifies to $[D_s]_{ij} = (1/V) \partial^2 \Omega/\partial q_i
\partial q_j\big|_{\vec{q}=\vec{0}}$ for systems with time-reversal symmetry (TRS) --- and assuming that the order parameter is always real, even for nonzero $\vec{q}$. The partial derivative is taken with all variables but $\vec{q}$ is kept constant, meaning only knowledge of the ground state is required (e.g., only $\Delta(\vec{q}=0)$ is needed, not $\Delta(\vec{q}\neq 0)$). This
simplified equation has been used for example to show that the
superfluid weight of isolated flat bands is proportional to the
quantum metric. A salient problem with that result, however, is that the quantum metric
depends on the positions of the orbitals $\{\vec{\delta_{\alpha}}\}$ through \Eq{eq.fourier}. On the
other hand, the superfluid weight is invariant under changes of $\{\vec{\delta_{\alpha}}\}$: this is immediately clear from the definition~\eqref{eq.sfw}, given
that the free energy does not depend on intra-unit cell
positions (when the hopping amplitudes $t_{i\alpha , j\beta}$ have been fixed constant). Using the terminology introduced in Ref.~[\onlinecite{Simon2020}],
the superfluid weight is geometry-independent while the quantum metric
is geometry-dependent.  The
source of this discrepancy is the assumption that all order
parameters are real even at nonzero $\vec{q}$. For a single-band
model, this assumption can always be made, because of the freedom in
the phase of the order parameter. However, for a multiband model, the
order parameters can have orbital-dependent phases, and cannot, in general, be
made simultaneously real by changing only the overall phase.

To understand how the problem arises, let us express ${\rm d}^2F/{\rm
  d}q_i{\rm d}q_j$ in terms of partial derivatives of the grand
canonical potential. For all the equations, we will fix the overall
phase of the order parameters by imposing reality and positivity on  a nonzero order parameter for one of the orbitals;  we choose it to be $\Delta_1(\vec{q})$. For simplicity, we will focus here on a
system with time reversal symmetry, which implies that
$\mu(\vec{q})=\mu(-\vec{q})$ and $\Delta_{\alpha}(\vec{q}) =
\Delta_{\alpha}^*(-\vec{q})$~\cite{Peotta2015}. Hence at
$\vec{q}=\vec{0}$, the
derivatives of the order parameters are purely imaginary  and ${\rm d}\mu/{\rm d}q_i\big|_{\vec{q}=\vec{0}}=0$. The general case
without TRS is treated in Appendix~\ref{app.sfw}. Using the chain
rule, the first derivative of the grand potential may be written as 
\begin{equation}
  \frac{{\rm d}\Omega}{{\rm d}q_i} = \frac{\partial \Omega}{\partial
    q_i} + \frac{\partial \Omega}{\partial 
    \mu} \frac{{\rm d}\mu}{{\rm d}q_i} + \sum_{\alpha} \frac{\partial
    \Omega}{\partial \Delta_{\alpha}^{I}} \frac{{\rm
      d}\Delta_{\alpha}^{I}}{{\rm 
      d}q_i} + \sum_{\alpha} \frac{\partial \Omega}{\partial
    \Delta_{\alpha}^{R}} \frac{{\rm d}\Delta_{\alpha}^{R}}{{\rm
      d}q_i},
  \label{eq.step_chain}
\end{equation}
where we have used the notation $\Delta_{\alpha}^{I}={\rm
  Im}(\Delta_{\alpha})$ and $\Delta_{\alpha}^{R}={\rm
  Re}(\Delta_{\alpha})$. Taking the total derivative of
Eq.~\eqref{eq.step_chain} with reference to $q_j$ and setting
$\vec{q}=\vec{0}$ yields 
\begin{align}
  \frac{{\rm d}^2F}{{\rm d}q_i{\rm d}q_j}
  \bigg|_{\vec{q}=\vec{0}} &=
  \frac{{\rm d}^2\Omega}{{\rm d}q_i{\rm d}q_j}
  \bigg|_{\vec{q}=\vec{0}} - \frac{\partial \Omega}{\partial
    \mu}\frac{{\rm d}^2\mu}{{\rm d}q_i{\rm
      d}q_j}\bigg|_{\vec{q}=\vec{0}} \\
  &= \frac{{\rm d}}{{\rm d}q_j} \frac{\partial
    \Omega}{\partial q_i}\bigg|_{\vec{q}=\vec{0}} \\
  &= \frac{\partial ^2\Omega}{\partial q_i\partial
    q_j}\bigg|_{\vec{q}=\vec{0}} +
  \sum_{\alpha}\frac{\partial^2\Omega}{\partial
    \Delta_{\alpha}^{I}\partial q_i}\frac{{\rm d}\Delta_{\alpha}^{\rm
      I}}{{\rm d}q_j}\bigg|_{\vec{q}=\vec{0}}. \label{eq.not_final}   
\end{align}
We have used that $\partial \Omega/\partial \Delta_{\alpha} = 0$ at all $\vec{q}$, which is equivalent to the gap equation, and that the total particle number $N=-\partial\Omega/\partial \mu$ is constant. Due to TRS, the derivatives of the order parameters are purely imaginary at $\vec{q}=\vec{0}$ and ${\rm d}\mu/{\rm d}q_i\big|_{\vec{q}=\vec{0}}=0$, which is why only the total derivatives of $\Delta_{\alpha}^I$ appear on the third line. 
Since $\partial \Omega/\partial
\Delta_{\alpha}=0$ holds at all $\vec{q}$, we have
\begin{equation}
  0 = \frac{{\rm d}}{{\rm d}q_i} \frac{\partial\Omega}{\partial
    \Delta_{\alpha}^I}\bigg|_{\vec{q}=\vec{0}} =
  \frac{\partial^2\Omega}{\partial q_i\partial
    \Delta_{\alpha}^I}\bigg|_{\vec{q}=\vec{0}} + \sum_{\beta}
  \frac{\partial^2\Omega}{\partial \Delta_{\alpha}^I\partial
    \Delta_{\beta}^I}\frac{{\rm d}\Delta_{\beta}^I}{{\rm
      d}q_i}\bigg|_{\vec{q}=\vec{0}} . \label{eq.sys}
\end{equation}
Using this identity, we can write Eq.~\eqref{eq.not_final} in a more
concise form
\begin{align}
 \frac{{\rm d}^2F}{{\rm d}q_i{\rm d}q_j}\bigg|_{\vec{q}=\vec{0}} &=
  \frac{\partial^2\Omega}{\partial q_i\partial
    q_j}\bigg|_{\vec{q}=\vec{0}} - ({\rm 
    d}_i\Delta^I)^{\rm T} \partial_{\Delta^{I}}^2\Omega ({\rm
    d}_j\Delta^I)\big|_{\vec{q}=\vec{0}},  \label{eq.final}\\ 
  {\rm d}_i\Delta^I &= \left( \frac{{\rm d}\Delta_2^I}{{\rm d}q_i},
  \ldots, \frac{{\rm d}\Delta_{n}^I}{{\rm d} q_i}
  \right)^{\rm T},\\
  \partial_{\Delta^{I}}^2\Omega &= \begin{pmatrix}
    \frac{\partial^2\Omega}{\partial \Delta_{2}^I\partial
      \Delta_{2}^I} & \ldots & \frac{\partial^2\Omega}{\partial
      \Delta_{2}^I\partial \Delta_{n}^I} \\
    \vdots & \ddots & \vdots \\
    \frac{\partial^2\Omega}{\partial \Delta_{n}^I\partial \Delta_2^I}
    & \ldots & \frac{\partial^2\Omega}{\partial \Delta_n^I\partial
      \Delta_n^I} 
  \end{pmatrix}.
\end{align}
The partial derivatives in $\partial_{\Delta^I}^2\Omega$ are taken by varying the involved order parameter while keeping all other variables constant.
The order parameter $\Delta_1$ does not appear in ${\rm
  d}_i\Delta^I$ and $\partial_{\Delta^I}^2\Omega$ because we assumed that $\Delta_1$ is
always taken real and positive. If the overall phase of the order
parameters is not fixed, an additional row and column containing the
derivatives involving $\Delta_1$ needs to be added to $\partial_{\Delta^I}^2\Omega$.

Clearly, ${\rm d}^2F/{\rm d}q_i{\rm d}q_j|_{\vec{q}=\vec{0}} =
\partial^2\Omega/\partial q_i \partial q_j|_{\vec{q}=\vec{0}}$ when
${\rm d}\Delta_{\alpha}^{I}/{\rm d}q_i|_{\vec{q}=\vec{0}}=0$. This holds if the order parameters are real also at
nonzero $\vec{q}$. It has been argued in previous literature that the
simplified equation $[D_s]_{ij}=\partial^2\Omega/\partial q_i\partial q_i\big|_{\vec{q}=\vec{0}}$
can be used in systems with TRS, as in such systems the order
parameters can be made real  with a transformation of the form
$c_{i\alpha}\to c_{i\alpha}
e^{i\theta_{i\alpha}(\vec{q})}$~\cite{Peotta2015}. Since
this transformation has no effect on the eigenvalues of $H_{\rm BdG}$ or
on the absolute values of the order parameters, the free energy
remains unchanged, and there is no effect on the superfluid
weight. However, the derivatives of the order parameters (the rightmost term in Eq.~\ref{eq.final}), and
$\partial^2 \Omega/\partial q_i\partial q_j$, are \textit{not} conserved 
under this transformation; they both change in such a way that the left-hand side of Eq.~\ref{eq.final} remains invariant. Therefore, when using
$[D_s]_{ij}=(1/V)\partial^2\Omega/\partial q_i\partial
q_j|_{\vec{q}=\vec{0}}$, it is  
crucial to compute the partial derivative \textit{after} the
transformation $c_{i\alpha}\to c_{i\alpha}
e^{i\theta_{i\alpha}(\vec{q})}$ is performed. In practice, one cannot assume
that this simplified equation holds without knowledge of the behavior
of the order parameters at nonzero $\vec{q}$ even in systems with
TRS. This fact was correctly pointed out in
Ref.~[\onlinecite{Chan2022}]. However, it was stated therein that the additional terms
are zero when the orbitals are equivalent. This is not generally the
case: the introduction of the vector $\vec{q}$ in the systems
typically breaks the very symmetry of the lattice which guaranteed
equal pairing at the orbitals, meaning that the order parameters at nonzero
$\vec{q}$ can differ by a phase even if they are equal at
$\vec{q}=\vec{0}$. 

It is straightforward to show that the additional terms in Eq.~\eqref{eq.final} are always negative for $i=j$. 
The matrix $\partial_{\Delta^I}^2\Omega$ is the Hessian matrix of the grand canonical potential, and since the order parameters are a
minimum of $\Omega$, it is positive semidefinite. It follows
immediately that $({\rm d}_i \Delta)^T \partial_{\Delta^I}^2\Omega ({\rm
  d}_i\Delta)\geq 0$, which means that $(1/V)\partial^2\Omega/\partial q_i^2|_{\vec{q}=\vec{0}}
\geq [D_s]_{ii}$. This implies that
$\partial^2\Omega/\partial q_i\partial q_j$ can predict values that are much larger than the correct superfluid weight, including the case of indicating a nonzero superfluid weight when it is in fact
vanishing.

The derivatives ${\rm d}\Delta_{\alpha}/{\rm d}q_i$ can be
computationally expensive to evaluate, as they seem to require solving
the gap equation at different nonzero $\vec{q}$. Remarkably, however, their
computation requires only knowledge of the ground state at
$\vec{q}=\vec{0}$. This method relies on the system of equations given
in Eq.~\eqref{eq.sys}, which can be written in matrix form as
\begin{align}
  &(\partial_{\Delta^I}^2\Omega){\rm d}_i\Delta^{I} = -\vec{b_i}, \label{eq.lin_sys_del}\\
  &\vec{b_i} = \left( \frac{\partial^2\Omega}{\partial q_i\partial
    \Delta_{2}^{I}},\ldots, \frac{\partial^2\Omega}{\partial
    q_i\partial \Delta_{n}^{I}} \right)^{\rm T}.
\end{align}
The derivatives of the order parameters are thus ${\rm
  d}_i\Delta^I = -(\partial_{\Delta^I}^2\Omega)^{-1}\vec{b_i}$, which involves only
partial derivatives of $\Omega$ and does not require knowledge of the state at nonzero $\vec{q}$. Note that if we had not fixed the
overall phase of the order parameters by choosing $\Delta_1$ real and positive, the gap equation would have an infinite number of solutions due to the freedom in this phase. In this case, the Hessian matrix would contain the terms related to partial derivatives w.r.t. $\Delta_1^I$ and would not be invertible.

\subsection{Superfluid weight from linear response: the conventional and geometric contributions}

In previous literature, the superfluid weight has been split into
so-called conventional and geometric parts~\cite{Peotta2015,Liang2017}. The conventional part is
the only component present in single-band models, and is related to
the derivatives of the band structure. It vanishes in the flat-band
limit. The geometric part is a purely multiband component which can be
nonzero even on flat bands. Expressions for these components have been
derived from linear response theory~\cite{Liang2017}, but without
accounting 
for the dependence of the order parameters on the vector potential. We
compute the full mean-field superfluid weight from linear response
theory in Appendix~\ref{app.lin}, and obtain
\begin{widetext}
\begin{equation}
  [D_s]_{ij} = \frac{1}{V}
  \sum_{\vec{k},ab}\frac{n_F(E_a)-n_F(E_b)}{E_b-E_a}\big[ 
    \bra{\psi_a}\partial_{i}\widetilde{H}_{\vec{k}} \ket{\psi_b}
    \bra{\psi_b} \partial_{j}\widetilde{H}_{\vec{k}} \ket{\psi_a}
    -\bra{\psi_a}(\partial_{i}\widetilde{H}_{\vec{k}}\gamma^z+\delta_{i}\Delta)
    \ket{\psi_b} \bra{\psi_b}
    (\partial_{j}\widetilde{H}_{\vec{k}}\gamma^z
    +\delta_{j}\Delta)  
    \ket{\psi_a} 
    \big], \label{eq.linrespresult}
    \end{equation}  
\end{widetext}
where
\begin{align}
  \partial_{i}\widetilde{H_{\vec{k}}} &= \begin{pmatrix}
    \frac{\partial H_{\vec{k}'}^{\up}}{\partial
      k_{i}'}\bigg|_{\vec{k'}=\vec{k}} & 0 \\
    0 & \frac{\partial (H_{\vec{k}'}^{\dn})^*}{\partial
      k_{i}'}\bigg|_{\vec{k'}=-\vec{k}}
  \end{pmatrix}, \nonumber \\
  \delta_{i}\Delta &= \begin{pmatrix}
    0 & \frac{{\rm d}\vec{\Delta}}{{\rm d}q_{i}} \\
    \frac{{\rm d}\vec{\Delta}^{\dag}}{{\rm d}q_{i}} & 0
  \end{pmatrix}.
\end{align}
Here, $\gamma^z = \sigma_z\otimes \mathbf{1}_{n\times n}$, where $\sigma_i$ are Pauli matrices and $\mathbf{1}_{n\times n}$ is the $n\times n$ identity matrix. The eigenvalues
and eigenvectors of $H_{\rm BdG}$ are $E_a$ and $\ket{\psi_a}$
respectively, and $n_{F}(E)$ is the Fermi-Dirac distribution at $E$. The prefactor in~(\ref{eq.linrespresult}) should be understood as $-\partial
n_F(E)/\partial E$ when $E_a=E_b$. This expression differs from the one given in~[\onlinecite{Liang2017}] by the addition of $\delta_i\Delta$ in the second term on the RHS of Eq.~\eqref{eq.linrespresult}, which accounts for the derivatives of the order parameters.

To separate the conventional and
geometric contributions, we write the eigenvectors in terms of the
Bloch functions $\ket{m_{\vec{k}}}_{\sigma}$: $\ket{\psi_a}=\sum_{m=1}^{n}
(w_{+,am} \ket{+}\otimes\ket{m_{\vec{k}}}_{\up}+w_{-,am}\ket{-}\otimes\ket{m^*_{-\vec{k}}}_{\dn})$, where
$\ket{m_{\vec{k}}}_{\up}$ is the eigenvector of $H_{\vec{k}}^{\up}$ with
eigenvalue $\epsilon_{\up,m,\vec{k}}$, $\ket{m^*_{-\vec{k}}}_{\dn}$ is the
eigenvector of $(H_{-\vec{k}}^{\dn})^*$ with eigenvalue
$\epsilon_{\dn,m,-\vec{k}}$, and $\ket{\pm}$ are the eigenvectors of $\sigma_z$ with eigenvalues $\pm 1$. Then
\begin{align}
  [D_{s,{\rm conv}}]_{\mu\nu} = \sum_{\vec{k}}\sum_{mn} C_{nn}^{mm}
  [j_{\mu}^{\up}&(\vec{k})]_{mm} [j_{\nu}^{\dn}(\vec{k})]_{nn}, \nonumber \\
  C_{pq}^{mn} = 4 \sum_{ab} \frac{n_F(E_a)-n_F(E_b)}{E_b-E_a}&
  w_{+,am}^*w_{+,bn} w_{-,bp}^* w_{-,aq}, \nonumber \\
  [j_{\mu}^{\sigma}(\vec{k})]_{mn} = \phantom{a}_{\sigma}\bra{m_{\vec{k}}}&
  \partial_{\mu} H_{\vec{k}}^{\sigma} \ket{n_{\vec{k}}}_{\sigma},
\end{align}
where $\partial_{\mu} = \partial/\partial_{k_{\mu}}$. The geometric
contribution is $D_{s,{\rm geom}} = D_{s} - D_{s,{\rm conv}}$.

The expression for the conventional contribution matches the one given in Ref.~[\onlinecite{Liang2017}], but the geometric contribution contains terms arising from the derivatives of the imaginary components of the order parameters. All the new additional terms in Eq.~\eqref{eq.final} (i.e., other than the partial derivative of the grand potential) are thus added to the geometric contribution, which is reasonable, as they can only be nonzero in multiband models. This split into conventional and geometric contributions is independent of the choice of orbital positions, and as we show below, the geometric part is related to the minimal quantum metric in isolated flat bands. These definitions are valid in a system with TRS, where the derivatives of the order parameters can be made purely imaginary at $\vec{q}=0$~\cite{Peotta2015}. In a system without TRS, there are additional terms arising from the derivatives of the real parts of the order parameters which can be nonzero even in a single-band system.

The superfluid weights derived from the free energy, Eq.~\ref{eq.final}, and by linear response, Eq.~\ref{eq.linrespresult} are equal, as shown in Appendix~\ref{app.equivalence}. We have verified numerically that both methods yield the same results in all examples studied in this article. 

\section{Quantum metric and isolated flat bands} \label{sec.qm}

The quantum metric of a set of bands $\mathcal{S}$ is the real part of the quantum geometric tensor
\begin{equation}
\mathcal{B}_{ij}(\vec{k}) = 2 \text{Tr }P(\vec{k}) \del_i P(\vec{k})\del_j P(\vec{k}) \label{eq.QGtensor}
\end{equation}
where $P(\vec{k}) = \sum_{m\in \mathcal{S}} \ket{m_{\vec{k}}}\bra{m_{\vec{k}}}$ is the projector into the Bloch states of the bands at $\vec{k}$. The quantum metric has been
previously related to the superfluid weight, most prominently in the
limit of isolated flat bands with TRS and where the pairing is uniform
in all orbitals where $\Delta_{\alpha}\neq 0$,
i.e. $\Delta_{\alpha}=\Delta$ for all $\Delta_{\alpha}\neq
0$~\cite{Peotta2015,Tovmasyan2016,Liang2017}. In such systems, the
superfluid weight is given by 
\begin{align}
\label{eq:DsQM}
  [D_{s}]_{ij}  &= \frac{4f(1-f)}{(2\pi)^{D-1}}|U|n_{\phi}\mathcal{M}_{ij},\\
  \mathcal{M}_{ij} &= \frac{1}{2\pi} \int_{\rm B.Z.} {\rm d}^2\vec{k} \, {\rm
    Re }(\mathcal{B}_{ij}(\vec{k})).
\end{align}
Here $f$ is the filling fraction of the band, $\mathcal{M}_{ij}$ is the quantum metric of the isolated flat band, $n_{\phi}^{-1}$
is the number of orbitals where pairing is nonzero and $D$ is the dimension of the system. This result is
derived from mean-field theory using the equality $[D_{s}]_{ij} =(1/V)
\partial^2\Omega/\partial q_i\partial q_j\big|_{\vec{q}=\vec{0}}$, or the equivalent linear
response equations~\cite{Peotta2015,Liang2017,Julku2016}. However, as
we have shown in 
Section~\ref{sec.sfw}, this equation is only accurate in special cases, even in
systems with TRS and uniform pairing. We will show here that, nevertheless, it is actually possible to derive a general connection between the superfluid weight and the quantum geometry, but the relevant quantity turns out to be the {\it
  minimal} quantum metric, i.e. the quantum metric with the lowest
possible trace over all possible orbital positions. 

As stated in Sec.~\ref{sec.fromfreeenergy}, $\partial^2\Omega/(\partial q_i)^2\big|_{\vec{q}=\vec{0}} \geq
{\rm d}^2F/({\rm d}q_i)^2\big|_{\vec{q}=\vec{0}}$ in presence of TRS. Without TRS, this inequality may not be true when ${\rm d}\mu/{\rm d}q_i\big|_{\vec{q}=\vec{0}}\neq 0$ (see Sec.~\ref{sec.no_trs}). When the inequality is saturated, the quantum
metric is directly related to the superfluid weight. Otherwise, it
gives an upper bound. We will first show that in systems with TRS,
there always exists a point where the inequality is saturated. The
property $\Delta(\vec{q}) = \Delta^*(-\vec{q})$ implies that
\begin{equation}
  \frac{{\rm d}\Delta_{\alpha}}{{\rm d}q_i}\bigg|_{\vec{q}=\vec{0}} =
  i\Delta_{\alpha} \frac{{\rm d}\theta_{\alpha}}{{\rm
      d}q_i}\bigg|_{\vec{q}=\vec{0}}, \label{eq.der_delta} 
\end{equation}
where $\theta_{\alpha}$ is the phase of the order parameter
$\Delta_{\alpha} = |\Delta_{\alpha}|e^{i\theta_{\alpha}}$. As before,
we fix $\theta_1=0$, with $\Delta_1$ a nonzero order parameter. 
It follows from Eq.~\eqref{eq.der_delta} that 
${\rm d}\Delta_{\alpha}/{\rm d}q_i = 0$ can only be nonzero if
${\rm d}\theta_{\alpha}/{\rm d}q_i=0$, or if
$\Delta_{\alpha}=0$ meaning there is no pairing in the orbital. 

Let us now assume that the order parameters for a choice of intra-unit-cell positions $\{\vec{\delta_{\alpha}}\}$ are
$|\Delta_{\alpha}(\vec{q})| e^{i\theta_{\alpha}(\vec{q})}$. The order
parameters in the same model for another choice of positions
$\{\vec{\delta_{\alpha}} + \vec{x}_\al\}$ are
$|\Delta_{\alpha}(\vec{q})|e^{i\widetilde{\theta_{\alpha}}(\vec{q})}$,
with $\widetilde{\theta_{\alpha}}(\vec{q}) = \theta_{\alpha}(\vec{q}) -
2\vec{q}\cdot \vec{x}_\al$ (see Appendix.~\ref{app.orb_pos}). Therefore
\begin{equation}
  \frac{{\rm d}\widetilde{\theta_{\alpha}}(\vec{q})}{{\rm d}q_i} =
  \frac{{\rm d}\theta_{\alpha}(\vec{q})}{{\rm d}q_i} -
  2 x^i_\al. 
\end{equation}
To set ${\rm d}\Delta_{\alpha}/{\rm d}q_i=0$ and guarantee that 
$[D_s]_{ij}=(1/V)\partial^2\Omega/\partial q_i\partial q_j\big|_{\vec{q}=\vec{0}}$, 
we can thus shift the orbital positions by
\begin{equation}
x^i_\al = \frac{1}{2}\frac{{\rm d} \theta_{\alpha}(\vec{q})}{{\rm
  d}q_i}\bigg|_{\vec{q}=\vec{0}}. \label{eq.positions}
\end{equation}
With the overall phase of the order parameters fixed, the order parameters are uniquely defined, and this shift is unique for all orbitals where $\Delta_{\alpha}\neq 0$. The resulting positions $\{\vec{\delta_{\alpha}+\vec{x_{\alpha}}}\}$ are independent of the particular initial choice of $\{\vec{\delta_{\alpha}}\}$ (see Appendix~\ref{app.unique}). If we had not fixed the overall phase, the positions where $[D_{s}]_{ij} =(1/V)
\partial^2\Omega/\partial q_i\partial q_j\big|_{\vec{q}=\vec{0}}$ would be unique up to an overall translation. 
The quantum metric computed for this appropriate set of positions is related to the superfluid weight directly. We find precise analogs of these results in the uniform pairing Hubbard models considered in Sec.~\ref{sec.Jonah}.

We have shown that positions $\{\vec{\delta_{\alpha}} + \vec{x}_\al\}$ where $D_s$ is related to the quantum metric exist, but solving them from Eq.~\eqref{eq.positions} requires knowledge of the derivative of the order parameters at some set of orbital positions $\{\vec{\delta_{\alpha}}\}$. This would still require solving the gap equation at a finite $\vec{q}$ to know which quantum metric is related to the superfluid weight. We will now show that it is possible to compute the correct quantum metric {\it without} solving the gap equation: it is the one with the smallest possible trace. 

As shown previously~\cite{Peotta2015}, $\partial^2\Omega/\partial q_i\partial q_j \propto \mathcal{M}_{ij}$,
and $\partial^2\Omega/\partial q_i^2 \geq {\rm d}^2F/{\rm
  d}q_i^2$. The result obtained from the quantum metric is thus always
an upper bound for the diagonal components of the superfluid weight,
and this upper bound is tight for the particular choice of positions
that makes the derivatives of the order parameters zero: this is thus a
minimum over all possible choices of orbital positions. For an
isolated flat band, the quantum metric with the \textit{smallest possible
integral of its diagonal components is thus proportional to the
superfluid weight}. Since all diagonal components are as small as
possible, this is the quantum metric with the smallest possible
trace.

The relationship between the superfluid weight and the quantum metric
has been used to derive lower bounds for the superfluid weight in flat
band systems. Our result shows that for such a lower bound to be
valid, it needs to be a lower bound for the quantum metric for any
choice of the orbital positions. The validity of some lower bounds
found in literature is discussed in Sec.~\ref{sec.prev}. 

\section{Example: superfluid weight, quantum metric, and orbital positions in the Lieb lattice} \label{sec.example}

\begin{figure}
  \centering
  \includegraphics[width=\columnwidth]{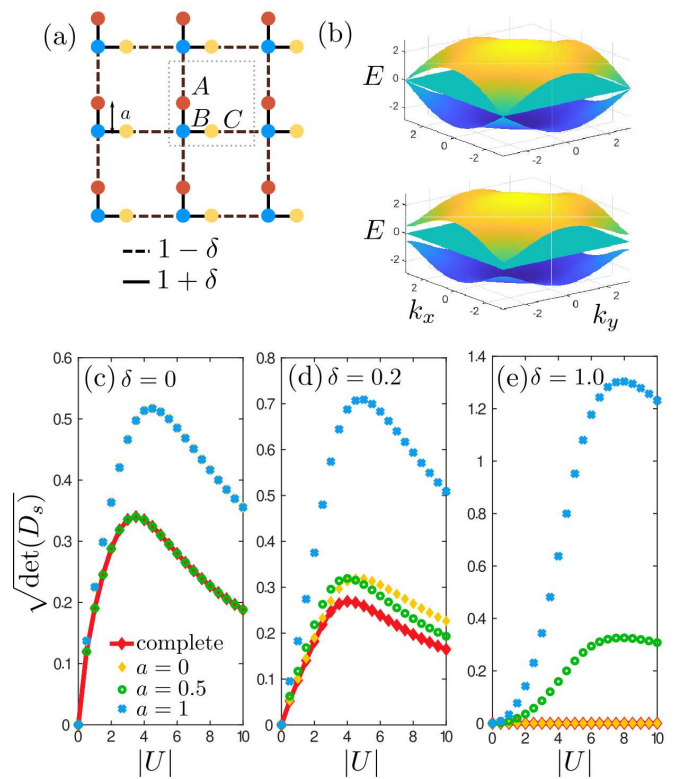}
  \caption{(a) Lieb lattice with staggered hopping amplitudes. The
      position of the orbitals in the unit cell is controlled by the
      parameter $a$. The typical Lieb lattice is $C_4$-symmetric corresponding to $a=\frac{1}{2}$,
      while $a=0$ and $a=1$ are equivalent to Fourier transformations
      where the positions of the orbitals are ignored with different
      choices of the unit cell. (b) Single-particle band structure at
      $\delta=0$ and $\delta=0.2$. The flat band is separated from the
      other bands by a band gap $E_{\rm gap} = \sqrt{8}\delta$. 
      (c-e) Superfluid weight $\sqrt{{\rm det}(D_s)}$ in the Lieb
      lattice computed with (red, "complete") and without (the other colors) the corrections for three
      different choices of intra-unit cell
      positions.} \label{fig.example}  
\end{figure}

To illustrate the importance of the additional terms of superfluid weight derived in Sections~\ref{sec.sfw} and~\ref{sec.qm}, and the role of orbital positions, we
study the superfluid weight in the Lieb lattice, shown in
Fig.~\ref{fig.example}a). This model has time-reversal symmetry and is invariant under the interchange of the $A$ and $C$ orbitals When $\delta=0$ and $a = \frac{1}{2}$, the Lieb lattice possess $C_4$ rotation symmetry, inversion symmetry, and reflection symmetry that interchanges the $A, C$ orbitals, and thus belongs to symmetry group $C_{4v}$.  Changing $\delta \neq 0$ or $a \neq \frac{1}{2}$ destroys the $C_4$ and inversion symmetries, but the mirror symmetries are preserved, thus reducing to symmetry group to $C_s$.  The flat band states reside solely on the $A$ and $C$ sites. The staggering of the hopping amplitudes is controlled by the parameter $\delta$, and introduces a band gap $E_{\rm gap} = \sqrt{8} \delta$, as shown in Fig.~\ref{fig.example}b.
We employ the parameter $a$ to control the distance between the $B$ site and the $A/C$ sites in a unit cell, and take the volume of a unit cell to be $1$.
We use the average inter-site hopping amplitude as our energy unit. The complete equation~(\ref{eq.final}) yields a result that is independent of the choice of orbital positions (see Fig.~\ref{fig.example}c-e), contrary to 
$(1/V)\partial^2\Omega/\partial q_i\partial q_j\big|_{\vec{q}=\vec{0}}$. In the
extreme case $\delta=1$, when the lattice is disconnected and can
clearly not support superconductivity, the correct superfluid weight
is zero. However, using $\partial^2\Omega/\partial q_i\partial q_j\big|_{\vec{q}=\vec{0}}$
can in fact give a nonzero and quite large superfluid weight.

At $\delta=0$, the simplified equation $[D_s]_{ij} = (1/V)\partial^2\Omega/\partial q_i\partial q_j\big|_{\vec{q}=\vec{0}}$ holds exactly when $a=\frac{1}{2}$ (see Fig. \ref{fig.example}),
which corresponds to the Lieb lattice with $C_{4v}$ symmetry when the convention given by Eq.~\eqref{eq.fourier} for the Fourier transformation is
used. This is explained by the equal hopping amplitudes in all directions: the
systems with $a=\frac{1}{2}-x$ and $a=\frac{1}{2}+x$ are identical up to an overall
rotation, and the additional terms are thus symmetric around $a=\frac{1}{2}$,
where the minimum of 
$\partial^2\Omega/\partial q_i\partial q_j$ occurs. Our proof in Sec.~\ref{sec.Jonah} generalizes this statement to all space groups. When
$\delta$ is increased and $C_4$ symmetry is broken, the orbital positions for which the relation
$[D_s]_{ij}=(1/V)\partial^2\Omega/\partial q_i\partial 
q_j\big|_{\vec{q}=\vec{0}}$ holds shifts continuously towards $a=0$. Importantly, there is a
wide parameter range where none of the choices $a=\frac{1}{2}$, $a=0$ or $a=1$
give the correct result when the derivatives of the order parameters
are ignored. When $a=0$ or $a=1$, the position of the $A, C$ orbitals is at the unit cell origin (where the $B$ orbitals are), and hence the Fourier transform Eq.~\ref{eq.fourier} becomes identical to the other convention Eq.~\ref{eq.fourier_no}. 

\begin{figure}
  \centering
  \includegraphics[width=\columnwidth]{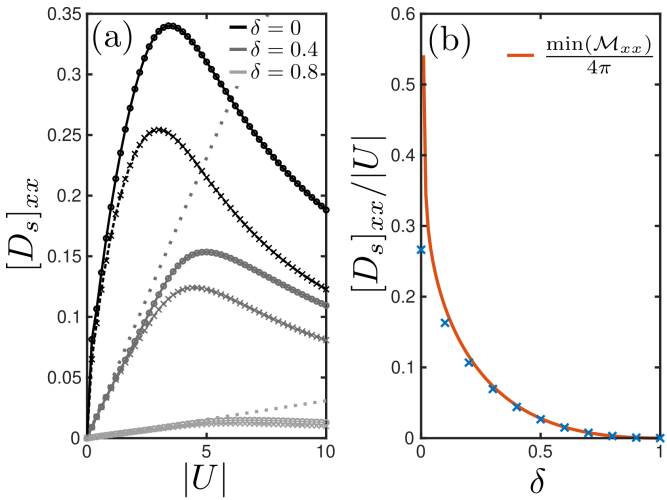}
  \caption{(a) Superfluid weight (circles) and geometric contribution
    (crosses) as a function of $|U|$ at different $\delta$ in the Lieb
  lattice. The dotted lines indicate the predictions from the minimal
  quantum metric. Only $[D_s]_{xx}$ is shown as the off-diagonal
  components of the superfluid weight tensor are very small for all
  parameters. (b) $[D_{s}]_{xx}/|U|$ at low interactions obtained from a
  linear fit (crosses) and prediction for the slope from the minimal
  quantum metric.} \label{fig.qm_sf}
\end{figure}

Finally, let us consider the role of the conventional and geometric parts of superfluidity in our example case. In an earlier study~\cite{Julku2016}, the quantum metric in the Lieb lattice has
been related to the superfluid weight. Note
that while only the $A$ and $C$ sites have equal pairing, the order
parameter on the $B$ sites is vanishing in the isolated band
limit, meaning the uniform pairing condition is fulfilled. As shown in Fig.~\ref{fig.qm_sf}a, the main contribution to
the superfluid weight at low interactions is the geometric part, and
the ratio $D_{\rm geom}/D_s$ approaches one in the isolated flat band
limit. This is expected as the conventional contribution should vanish
on a perfectly flat band. The prediction from the minimal quantum metric, shown in Fig.~\ref{fig.qm_sf}b), is increasingly accurate with growing $\delta$. 

\section{Cooper Pair Mass Beyond Mean Field} \label{sec.Jonah}

It has been shown that the two-body problem in a flat band gives for the bound pair a finite effective mass that is governed by quantum geometry~\cite{Torma2018,Iskin2021}. For uniform pairing, the inverse effective mass can be approximately related to the quantum metric. Thus pairs can move while single particles cannot, meaning that the qualitative picture given by mean-field superfluid weight calculations is already apparent at the two-body level. Here we calculate the Cooper pair mass in a full many-body treatment and without a mean-field approximation. The mass is obtained from the spectrum of pair excitations of the ground state. It shows dependence on quantum geometry and allows relating the proper choice of quantum metric discussed above to the system symmetries.     

We consider a family of positive semi-definite, $D$-dimensional, attractive Hubbard models first introduced by Ref.~[\onlinecite{Tovmasyan2016}] where the electron kinetic energy term has $N_f$ perfectly flat zero-energy bands fulfilling a condition where the single-particle projectors $P(\vec{k})$ (see \Eq{eq.QGtensor}) obey
\bea
\label{eq:upc}
\int \frac{d^Dk}{(2\pi)^D} P_{\al \al}(\vec{k}) = n_\phi N_f \equiv \eps
\eea
for all orbitals $\alpha =1,\dots,n_\phi^{-1}$ where the pairing is nonzero. The condition (\ref{eq:upc}) leads to the pairing gaps on different orbitals being the same, therefore it is also referred to as the uniform pairing condition. We neglect the spin label, assuming that the model has time-reversal symmetry which relates the two projectors: $P_\uparrow(\vec{k})=P^*_\downarrow(-\vec{k})\equiv P(\vec{k})$. Upon projecting the many-body operators into the $N_f$ flat bands, the kinetic energy vanishes and the Hamiltonian is given by the interaction term
\bea
H_{U} &= -|U| \sum_{i\al} \bar{n}_{i\al,\u} \bar{n}_{i\al,\d} + \frac{n_\phi N_f}{2} |U| \bar{N} , \\
\eea
where $\bar{n}_{i,\al ,\sigma}$ is the projected density operator in orbital $\al$ and spin $\sigma$, $\bar{N}$ is the projected total density operator. Ref.~[\onlinecite{Tovmasyan2016}] demonstrated that $H_{U}$ possesses $\eta$-pairing groundstates, that is, states with all particles paired. In forthcoming work Ref.~[\onlinecite{upcoming}], we show that the Cooper pair excitations on top of these groundstates are exactly solvable thanks to the uniform pairing condition, and we are able to calculate their effective mass exactly. 

The Cooper pair excitations are governed by the following single-particle Hamiltonian:
\bea
\label{eq:hp}
h_{\al \be}(\vec{q})  &= \int \frac{d^Dk}{(2\pi)^D} P_{\al \be}(\vec{q}+\vec{k}) P_{\be \al}(\vec{k}) \ . \\
\eea
We denote the eigenvalues of $h(\vec{q})$ as $\eps_\mu(\vec{q})$, where $\mu = 0,\dots, n_{\phi}^{-1}-1$. The many-body energy of the lowest lying Cooper pair is $|U|(\eps - \eps_0(\vec{q}))$, where $\eps_0(\vec{q})$ is the \textit{largest} eigenvalue of $h(\vec{q})$.

We now show that $\eps_\mu(\vec{q})$, and hence the Cooper pair spectrum, is invariant under a redefinition of the orbital locations $\vec{\delta}_\al \to \vec{\delta}_\al + \vec{x}_\al $ (leaving the hopping elements invariant). This must be the case physically because the choice of $\vec{x}_\al$ is just a convention for the Fourier transform. Since the redefinition means $P_{\al \be}(\vec{k}) \to e^{-i \vec{k} \cdot (\vec{x}_\al-\vec{x}_\be)} P_{\al \be}(\vec{k})$, we see that $h_{\al \be}(\vec{q})$ transforms under a redefinition of the orbitals as
\bea
h_{\al \be}(\vec{q}) &\to e^{-i \vec{q} \cdot (\vec{x}_\al-\vec{x}_\be)} \int \frac{{\rm d}^Dk}{(2\pi)^D} P_{\al \be}(\vec{q}+\vec{k}) P_{\be \al}(\vec{k}) \\
&= [V^\dag_{\vec{x}}(\vec{q}) h(\vec{q}) V_{\vec{x}}(\vec{q})]_{\al \be} 
\eea
where we defined the diagonal unitary matrix $[V_{\vec{x}}(\vec{k})]_{\al \be} = e^{i \vec{k} \cdot \vec{x}_\al} \delta_{\al \be}$. We see explicitly that, although $h(\vec{q})$ is not invariant, its spectrum is. 

 The effective Cooper pair mass is given by 
\bea
\null [m^{-1}]_{ij} = \left. -|U| \frac{{\rm d}^2 \eps_0(\vec{q})}{{\rm d}q_i{\rm d}q_j} \right|_{\vec{q}=\vec{0}}
\eea
which is computed from the spectrum of $h(\vec{q})$ and thus is manifestly invariant. Using perturbation theory, $\eps_0(\vec{q})$ can be easily calculated to second order in $\vec{q}$. At zeroth order $\eps_0(0) = \eps$, which corresponds to the constant eigenvector $u_0^\al = \sqrt{n_\phi}$. The first order correction vanishes (showing the Cooper pair is stable), and we calculate two contributions at second order in $q_i$:
\bea
\label{eq:expansion}
\eps(\vec{q}) &= \eps + \sum_{\mu =1}^{n-1} \frac{| u_\mu^\dag (\vec{q} \cdot \pmb{\nabla} h) u_0|^2}{\eps - \eps_\mu(0)} \\
&+ \frac{1}{2} q_i q_j \int \frac{d^Dk}{(2\pi)^D}\sum_{\al \be} u_0^\al \del_{ij}  P_{\al \be}(\vec{k}) P_{\be \al}(\vec{k}) u_0^\be ,
\eea
noting that $\eps_\mu(0) < \eps$ are the eigenvalues of $h(0)$, so the first line is non-negative, and where $\pmb{\nabla} h$ is the gradient of $h$ evaluated at ${\vec q} = 0$. After integration by parts, the integral in the second line yields
\bea
\label{eq:intbyparts}
&n_\phi \sum_{\al \be} \int \frac{d^Dk}{(2\pi)^D}  \del_{ij} P_{\al \be} P_{\be \al} \\
&=  -n_\phi \int \frac{d^Dk}{(2\pi)^D}  \text{Tr } \del_i P \del_j P = - \frac{n_\phi}{(2\pi)^{D-1}} \mathcal{M}_{ij},
\eea
which is proportional  the quantum metric integrated over the Brillouin zone, i.e.~$ \mathcal{M}_{ij}$ defined in Eq.(\ref{eq:DsQM}) (as $\text{Tr }P \{\del_i P, \del_j P\} = \text{Tr }\del_i P \del_j P$). Hence \Eq{eq:intbyparts} is negative semi-definite.

It is important to note that $\pmb{\nabla} h$ is not invariant under the choice of $\vec{x}_\al$, transforming as
\bea
\pmb{\nabla} h_{\al\be} \to \pmb{\nabla} h_{\al\be} - i (\vec{x}_\al - \vec{x}_\be) h_{\al \be}(0) \ . \\
\eea
Nevertheless, it is possible to show that, up to a choice of origin, there is a unique choice of $\vec{x}_\al$ where $\pmb{\nabla} h u_0 = 0$ and the quantum metric is the sole contributor to the effective mass. Note that the $O(p^2)$ term in the first line of \Eq{eq:expansion} competes with $ -\mathcal{M}_{ij}$ in \Eq{eq:intbyparts} because it is opposite in sign. Thus the choice of $\vec{x}_\al$ where only the quantum metric is nonzero corresponds to the orbital positions of the minimal quantum metric. 

A calculation using the uniform pairing condition results in an explicit form for the orbital shifts that make the quantum metric the sole contribution for the effective mass, namely 
\bea
\label{eq:xalso}
 (\eps-h(0)) \vec{x}_\al &= -i[ \pmb{\nabla}h u_0]_\al .  \ \\
\eea
This equation has a unique solution up to the overall choice of origin because $\eps-h(0)$ has a single zero-eigenvalue corresponding to the uniform eigenvector $u_0$.
With the the orbital shifts $\vec{x}_\al$ given by Eq.~(\ref{eq:xalso}), the effective mass becomes
\bea
\label{eq:mquanummetric}
[m^{-1}]_{ij} = \frac{n_\phi}{(2\pi)^{D-1}} |U| \mathcal{M}_{ij} .
\eea
Comparing this equation to \Eq{eq:DsQM}, we find exact agreement with the mean field superfluid weight up to an overall factor of $4f(1-f)$,which is the Cooper pair density.

We now improve upon \Eq{eq:mquanummetric} in two ways. First we find that $\vec{x}_\al$ obey the space group symmetries $g \in G$ of the Hamiltonian when the symmetric choice of Fourier convention (\Eq{eq.fourier}) is used. In other words, when the symmetry-preserving positions of the orbitals are used, their deviations $\vec{x}_\al$ also obey the space group symmetries. In many cases, this is tantamount to a proof that $\vec{x}_\al = 0$, meaning that the quantum metric is the {\it minimal} quantum metric, and is the Cooper pair mass. For instance, at $\delta = 0$ in the Lieb lattice with $a=\frac{1}{2}$, the $A$ and $C$ orbitals are related by $C_4$ symmetry and are invariant under $C_2$. There is no way to deform these orbitals off the positions $a=\frac{1}{2}$ without breaking $C_2$. Thus $\vec{x}_\al = 0$, thereby explaining why $a = \frac{1}{2}$ is the correct choice to evaluate the minimal quantum metric in Fig. \ref{fig.example}. By a similar argument, all orbitals at fixed high-symmetry positions necessarily have $\vec{x}_\al = \vec{0}$ because they are pinned by symmetries. In these cases, the minimal quantum metric is guaranteed to be the one computed using the physical positions in \Eq{eq.fourier}.

 Secondly, we now propose a simple generalization of the uniform pairing condition that guarantees $\vec{x}_\al = 0$. We define the quantity
\bea
\varepsilon_\al(\vec{q}) = \int \frac{d^Dk}{(2\pi)^D} [P(\vec{k}+\vec{q}) P(\vec{k})]_{\al \al} 
\eea
which at $\vec{q}=0$ yields $\eps_\al = n_\phi N_f$, the uniform pairing condition in \Eq{eq:upc}. It is then direct to check that 
\bea
\label{eq:genUPC}
\varepsilon_\al(0) &= n_\phi N_f, \quad &\text{(uniform pairing condition)} \\
\del_i \varepsilon_\al(0) &= 0 , \quad &\text{(minimal metric condition)},
\eea
the latter condition being the many-body analogue of \Eq{eq.positions}, in that its solution sets the quantum metric to be minimal. 

These results directly parallel those given by mean field theory in the above sections. We have shown that the Cooper pair effective mass is independent of the Fourier convention for the orbital positions. Furthermore, there exists a choice of orbital positions where the effective mass is determined by the quantum metric alone and at these positions the quantum metric is minimal. Under the the uniform pairing condition, we provide an explicit formula for these positions in \Eq{eq:xalso}, to be compared to \Eq{eq.positions}. The inclusion of crystalline symmetries constrains the positions: if the orbital positions are pinned by the symmetries, then the quantum metric evaluated for those positions must be minimal. Lastly, we established a generalization of the uniform pairing condition in \Eq{eq:genUPC} to determine when the quantum metric is minimal.

\section{Non-isolated flat bands} \label{sec.ni}

The relationship of the minimal quantum metric and the superfluid
weight indicates that the BKT transition temperature could be
increased in systems with a high quantum metric. However, this is only
valid in the isolated flat band limit. The quantum metric typically
diverges when the band gap closes, but this is not an indication
that the superfluid weight diverges. The superfluid weight is proportional to $|U|\mathcal{M}_{ij}$ only when the flat band is isolated, which requires that the interaction strength is small compared to the band gap (otherwise pairing would involve higher bands). Therefore, when the band gap shrinks, the largest $|U|$ for which the quantum metric is proportional to $D_s$ decreases accordingly. The very large quantum metric that can be achieved with a small band gap is thus only relevant at very low interactions, where $[D_s]_{ij}\propto |U|\mathcal{M}_{ij}$ remains small. The divergence of the quantum metric is an indication that the contributions from the other bands are
important at low $|U|$, and reduce the superfluid weight compared to the isolated
flat band result. In the Lieb lattice, those contributions have been shown to curtail the divergence and lead to a finite superfluid weight~\cite{Julku2016}. An interesting question when searching for systems
with high $T_{\rm BKT}$ is whether the critical temperature can still
be large in the non-isolated band limit even though the contributions
from dispersive bands are prominent. In repulsive models, a flat band near the Fermi surface has been predicted to be beneficial~\cite{Aoki2020,Kuroki2005,Kobayashi2016,Matsumoto2018}. In attractive models, previous mean-field studies have indicated that the superfluid weight has a non-linear dependence on the interaction strength for non-isolated flat bands~\cite{Julku2016,Wu2021,Iskin2019}, but the additional terms we find in this work have not been taken into account.
In this section, we show by continuously tuning the band gap that the
superfluid weight and $T_{\rm BKT}$ can actually be maximal when there is a band
touching. Furthermore, we study its dependence on different types
of band touchings.  We supplement our analysis of band touching points by employing a $S$-matrix construction~\cite{Calugaru2021} to analyze bipartite lattices with band touching points.   

\subsection{Effect of closing the band gap}

\begin{figure}
  \centering
  \includegraphics[width=\columnwidth]{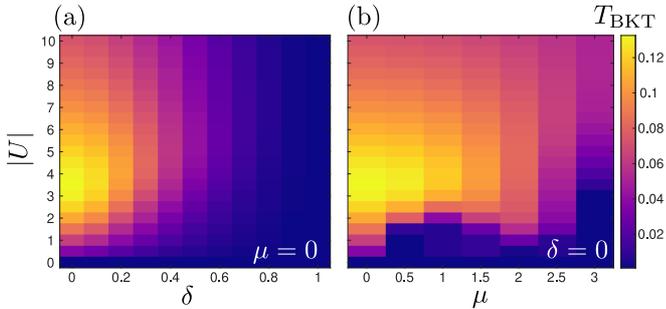}
  \caption{BKT temperature in the Lieb lattice (a) as a function of the
    hopping staggering $\delta$ and the interaction $|U|$, and (b) as a function of
    the chemical potential $\mu$ and the interaction.}\label{fig.ni_fb}
\end{figure}

As shown in Fig.~\ref{fig.qm_sf}, the superfluid weight in the Lieb lattice increases
monotonically when $\delta$ is decreased, and reaches its maximum when
$\delta=0$ for all interactions. At high interactions, the superfluid
weight decays as $\propto 1/|U|$, which is a well-known behavior
related to the formation of bound pairs in the BEC limit of the
BCS-BEC crossover~\cite{Iskin2019,Orso2021}. At low interactions,
$D_s\propto |U|$ when the flat band is isolated. This linear behavior is visible in an increasingly wide range of interactions when $\delta$ is increased. For $\delta=0$, when there is no band gap, the behavior
is no longer exactly linear, which is consistent with previous
literature such as~[\onlinecite{Julku2016}], and~[\onlinecite{Iskin2019,Wu2021}] where it has been found that $D_s\propto |U|{\rm ln}(C/|U|)$, with $C$ a constant. 

The superfluid weight at zero temperature is an upper bound for the
BKT temperature. However, it does not give the full picture: for
instance, the zero-temperature superfluid weight in a dispersive band
will typically be non-zero in the $U\to 0$ limit whereas it vanishes
in a flat band. At $T=0$, the superfluid weight will thus typically be
smaller in a flat band than a dispersive band for small interactions,
even though the BKT temperature is usually larger on the flat band
(see Fig.~\ref{fig.comp}). In this section, we solve the BKT
temperature from the universal
relation~\cite{Berezinsky1971,Kosterlitz1973,Nelson1977}
\begin{equation}
  T_{\rm BKT} = \frac{\pi}{8} \sqrt{{\rm det}(D_s(T_{\rm BKT}))}.
\end{equation}

As is shown in Fig.~\ref{fig.ni_fb}a, $T_{\rm BKT}$ mirrors the
behavior of the superfluid weight and increases monotonically with
$\delta$ for all considered interactions. The largest BKT temperature
occurs around interaction $U\approx -3.5$ with no hopping staggering so the flat band is not isolated. Moreover, the highest critical
temperature as a function of $\mu$ is found for the half-filled flat
band, showing that in this model, the highest possible critical
temperature is achieved in the flat band when it is not
isolated. Hence, the isolated flat band limit is not necessary
to reach a high $T_{BKT}$, and a band touching could actually be
beneficial for superconductivity. It is important to remember also that the flat band combined with a band touching yields a higher $T_c$ than a usual dispersive band (e.g., square lattice), for small interactions $|U|$, see Fig.~\ref{fig.comp}.

\subsection{Comparison of linear and quadratic band touchings} \label{sec.ltq}

\begin{figure*}
  \centering
  \includegraphics[width=\textwidth]{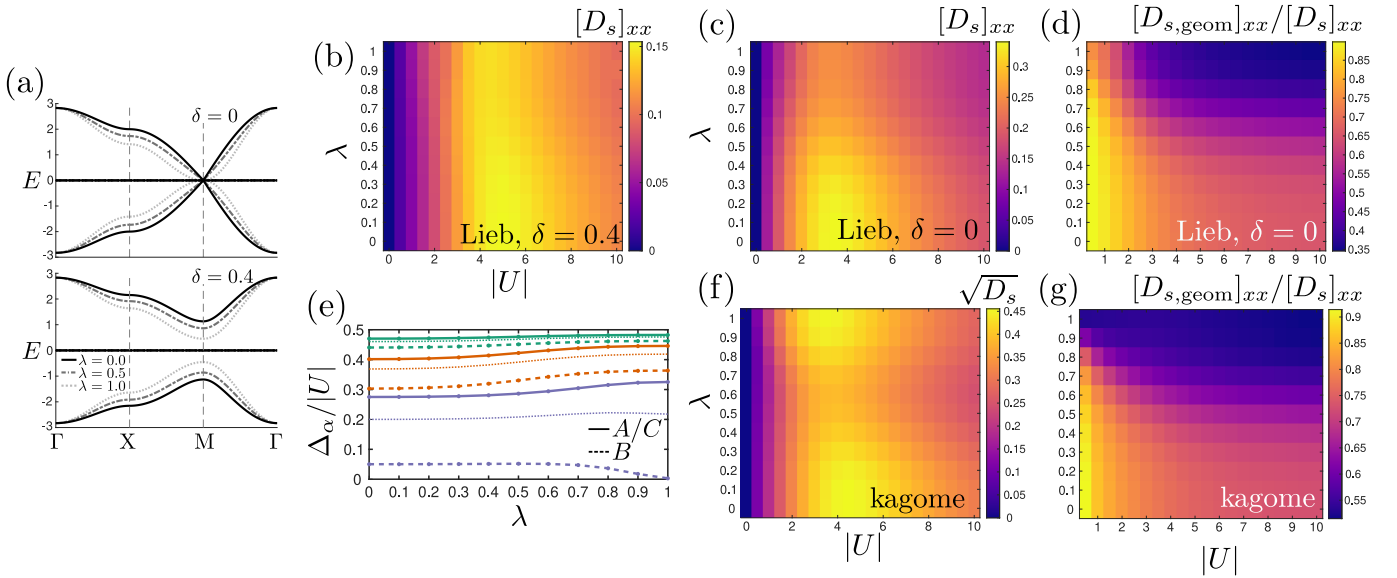}
  \caption{(a) Band structure of the tunable Lieb model for different
    values of $\lambda$, at $\delta=0$, i.e. in the presence of a band touching, and at $\delta=0.4$. The band touching can be tuned from linear ($\lambda=0$) to quadratic ($\lambda=1$) at $\delta=0$. At $\delta=0.4$, the dispersive bands are modified without changing the quantum metric of the flat band. (b) Superfluid weight $[D_s]_{xx}$ for
    $\delta=0.4$ in the Lieb model, when the flat band is separated
    from the other bands by a gap. 
  (c) Superfluid weight $[D_s]_{xx}$ 
    and (d) ratio $[D_{s,{\rm geom}}]_{xx}/[D_s]_{xx}$ in the tunable Lieb
    lattice. The off-diagonal components of the superfluid weight are
    zero.
    (e) Order parameters $\Delta_{\alpha}/U$ in the tunable Lieb lattice as a function of $\lambda$ at interaction strengths $U=-1$ (blue), $U=-4$ (orange) and $U=-8$ (blue). The order parameters in the $A/C$ orbital (full line) are always equal, and are larger than the order parameter in the $B$ orbital (dashed line). The dotted line shows the average of all order parameters.
  (f) $\sqrt{{\rm det}(D_s)}$
    and (g) $[D_{s,{\rm geom}}]_{xx}/[D_s]_{xx}$ in the tunable kagome
    model. In this case, the off-diagonal components are not always
    zero. A similar behavior of the ratio $[D_{s,{\rm
          geom}}]_{ij}/[D_s]_{ij}$ is observed for all components. 
    }
  \label{fig.ltq_res}
\end{figure*}

To study the effect of different types of band touchings on the
superfluid weight, we use the method developed in Ref.~[\onlinecite{Graf2021}] to
construct flat band models that can be continuously tuned from a
linear to a quadratic band touching. The method is based on building
two Hamiltonians $H_{\rm lin}$ and $H_{\rm quad}$ that feature a flat
band with a linear and quadratic band touching respectively, and for
which the flat band has exactly the same Bloch functions. Then the band
touching can be continuously tuned in the total Hamiltonian $H =
(\lambda H_{\rm quad} + (1-\lambda) H_{\rm lin})/C$ without affecting the
energy or the Bloch functions of the flat band. We study two such
models, constructed on a Lieb and kagome geometry. The tight-binding
parameters are given in Appendix~\ref{app.ltq}. These models both have a flat band at
$E=0$, and we pick $C$ so that the total width of
the band structure is independent of $\lambda$. Our energy unit is the average inter-site hopping strength of the $\lambda=0$ lattice model.  The band structure for
the tunable Lieb model is shown for three values of $\lambda$ in
Fig.~\ref{fig.ltq_res}a. The Lieb model is
constructed so that the band gap can be tuned with the staggering
parameter $\delta$.

In the Lieb model, when $\delta$ is nonzero, the superfluid weight at
low interactions becomes independent of $\lambda$ (see
Fig.~\ref{fig.ltq_res}b). This is expected,
as in the isolated band limit the superfluid weight is determined by
quantum geometry, and the flat band has the same quantum metric
for all $\lambda$. The range of interactions where $D_s$ is
independent of the parameter $\lambda$ grows with $\delta$, as the
band gap becomes larger and the isolated band limit is valid up to
larger $|U|$. At intermediate interactions, the limit $\lambda=0$,
corresponding to a linear touching, has a more pronounced
maximum. When the band gap is closed, differences when varying $\lambda$ occur already at
vanishingly small interactions, as shown in Fig.~\ref{fig.ltq_res}c-d). The superfluid weight is smaller
overall for the quadratic band touching $\lambda=1$. Moreover, the
ratio $D_{\rm geom}/D_s$ is much smaller for the quadratic than the
linear band touching. It is interesting to note that the superfluid weight behaves differently from the mean-field order parameters, shown in Fig.~\ref{fig.ltq_res}e. The order parameters at the $A/C$ sites are larger in the quadratic model than in the linear model for all interactions we consider. We show in Sec.~\ref{sec.mf_gap} that this is expected to hold in bipartite lattices with uniform pairing. However, even though the pairing is stronger in the quadratic model, the superfluid weight is lower, which is the opposite of what would be expected for an isolated flat band~\cite{Peotta2015,Liang2017} where the superfluid weight is proportional to the pairing gap.

For the kagome model, which does not feature a band gap, a similar
behavior of the geometric part of the superfluid weight can be
observed (see Fig.~\ref{fig.ltq_res}e-f)): its contribution is much more prominent for the linear band
touching than the quadratic one. The maximum of $D_s$ is also slightly
more pronounced in the linear model than the quadratic one, although
the superfluid weight is larger in the quadratic model at
small interactions.

The geometry of the flat band therefore does not give the full picture
in the non-isolated band limit: even though the Bloch functions of the
flat band are always the same when varying $\lambda$ which controls the type of band touching, the superfluid weight differs. This means that the behavior of the
superfluid weight is dependent on the nature of the band touching. 

\subsection{Band touching points from the $S$-matrix construction} \label{sec.mf_gap}

The mean field behavior of the pairing gap in general lattices, with both isolated and non-isolated flat bands, can be understood using the $S$-matrix construction of Ref.~[\onlinecite{Calugaru2021}]. This provides a description of the effect of band touchings on the pairing gap that is more general than given by the specific models considered above, and allows for an analytic solution in the mean field, yielding general results for quantities such as the pairing gap.  The power of this approach is made evident as it yields self-consistent gap equations independent of the wavefunctions, allowing for an analysis of pairing strength as a function of the lattice parameters and dispersion.

The $S$-matrix construction employs a bipartite lattice with two unequal sublattices $L, \tL$, with the difference between the number of orbitals per unit cell $N_L - N_\tL = N_f$ being the number of flat bands.  Band touching points can be enforced in the model via irrep analysis of the symmetries \cite{Calugaru2021}.
The bipartite Hamiltonian in such models reads
\begin{align}
H_\kk = \begin{bmatrix}
0 & S_\kk^\dagger \\
S_\kk & 0
\end{bmatrix},
\label{}
\end{align} where $S_\kk^\dagger$ is an $N_\tL \times N_L$ rectangular matrix encoding the hopping between the two sublattices. These $S$-matrix Hamiltonians can be realized in actual physical materials \cite{Regnault2021}. The energies come in $\pm \epsilon_{\kk,m}$ pairs, where $\epsilon$ are the singular values of $S_\kk$. Because $S_\kk^\dagger$ maps $C^{N_L}$ to $C^{N_\tL}$, there are at least $N_L- N_\tL$ vectors in the null space of $S_\kk^\dagger$; these form the flat bands.  One can introduce a quadratic Hamiltonian 
\begin{align}
    H_\text{quad} = H_\kk \begin{bmatrix}
    I_{\tL \times \tL} & 0 \\
    0 & -I_{L \times L}
    \end{bmatrix} H_\kk
\end{align} which has eigenvalues $\pm \epsilon_{\kk,m}^2, 0$, and preserves the flat band wavefunctions.  In the case of the Lieb lattice, this Hamiltonian is precisely the same as the Hamiltonian with quadratic band touching points studied in Sec.~\ref{sec.ltq}, obtained using the technique from  Ref.~[\onlinecite{Graf2021}] (see Appendix~\ref{app.ltq} for the tight-binding parameters of the model).  

By adding attractive on-site interactions and assuming that the pairing is uniform within each sublattice, that is, there are two gaps $\Delta_L$ and $\Delta_\tL$ depending on the sublattice, we find the following self-consistent gap equations at $T=0$ for $H_\kk$:

\begin{align}
N_L \Delta_L &= \frac{|U| N_\tL}{2} f(\Delta) + \frac{|U|(N_L - N_\tL)}{2}\\
N_\tL \Delta_\tL &= \frac{|U| N_\tL}{2} f(\Delta) ,
\label{eq:sysMF}
\end{align}  where $\Delta = \dfrac{1}{2}(\Delta_L + \Delta_\tL)$ and 
\begin{align}
    f(\Delta) &= \dfrac{1}{N_\tL} \sum_{ m =1}^{N_\tL} \int \frac{d^Dk}{(2\pi)^D} \dfrac{\Delta}{ \sqrt{\Delta^2 + \epsilon_{\kk,m}^2}} . \nonumber \\
\end{align} Here the sum is over the $N_\tL$ dispersive bands.  The function $f$ ranges from $0$ for a perfectly flat band at zero kinetic energy, to $1$ for a gapped band at very large kinetic energy, and is a monotonically increasing function of $\Delta$. \Eq{eq:sysMF} always has a solution, and obeys the following properties:
\begin{align}
    N_L \Delta_L - N_\tL \Delta_\tL &= \frac{|U|(N_L - N_\tL)}{2}, \label{eq:weighted_diff_linear}\\
    0 < \Delta_\tL &< \Delta_L < \frac{|U|}{2}, \\
    \frac{N_L - N_\tL}{4} &< \frac{\Delta}{|U|} < \frac{1}{2}.
    \label{eq:linear}
\end{align} The first equality generalizes the result found in the Lieb lattice by Ref.~[\onlinecite{Julku2016}], as it now applies to any bipartite lattice with uniform pairing within each sublattice, and agrees with our numerical calculations of the pairing gaps.  The dispersion does not need to be gapless for this equality to hold; only the bipartite nature of the underlying lattice is required.  These relations are proved in Appendix~\ref{app.gap}.  Regardless of the form of the bipartite lattice, even in the absence of a band touching, we have the result that the pairing strength on the larger sublattice $\Delta_L$ is always larger than the pairing on the smaller sublattice $\Delta_\tL$, due to the fact that the flat bands greatly enhance the pairing for the sublattice $L$ (see Appendix~\ref{app.gap}), and both $\Delta_L, \Delta_\tL$ are bounded by quantities depending on the number of flat and dispersive bands.

Though the exact details of $f(\Delta)$ depend on the dispersion of the kinetic energy, the fact that it is bounded suggests that most of the gap strength comes from the flat band contribution which is universal.  To maximize the strength of the pairing $\Delta_L$, we note that the self-consistent equation for $\Delta_L$ depends only on the ratio of the number of bands of the sublattices $r = \frac{N_\tL}{N_L}$.  This is saturated as $r \rightarrow 0$: thus, even in the presence of band touching points, more flat bands per total bands enhances the superconducting gap at $T=0$. If the dispersive bands are gapped from the flat bands, with the band gap $ \gg |U|$, $f(\Delta) \rightarrow 0$.  Thus, we approach the limit discussed in Ref.~[\onlinecite{Peotta2015}], where one may project the Hamiltonian into the flat bands and obtain an exactly solvable BCS ground state.  

The quadratic band touching point, i.e.~the case of $H_\text{quad}$, has a different set of self-consistent gap equations (see Appendix~\ref{app.gap}), due to the fact that the dispersive bands have different wavefunctions (though the flat band wavefunctions remain the same). The self-consistent equations still always possess a solution so long as flat bands exist.  An analysis shows that the weighted difference reads
\begin{align}
N_L \Delta_L - N_\tL \Delta_\tL &= \frac{|U| N_\tL}{2} (f(\Delta_L) - f(\Delta_\tL)) \nonumber \\
&+ \dfrac{|U|(N_L - N_\tL)}{2},
\label{}
\end{align} which increases relative to \Eq{eq:weighted_diff_linear} so long as $\Delta_L > \Delta_\tL$. We prove that there always exists a solution of the gap equations with this property (see Appendix~\ref{app.gap} for more details). 

To make further statements about the pairing gap $\Delta$, we compare $f(\Delta)$ for a quadratic dispersion versus a linear dispersion.  In general, a higher density of states of the kinetic energy close to zero energy will raise $f(\Delta)$, thereby raising $\Delta$.  Thus, we expect the quadratic band touching will have a stronger pairing gap than the linear band touching. It is interesting to compare this to our numerical results for different band touchings in the Lieb lattice (Fig.~\ref{fig.ltq_res}) where the quadratic band touching does not give highest value for the superfluid weight. The pairing gap, on the other hand, is larger in the quadratic model, in agreement with our prediction. The pairing gap $\Delta_L$ is influenced by the density of states, which indeed is larger for quadratic dispersion than a linear dispersion. However, the superfluid weight depends also on quantum geometry which affects the ability of Cooper pairs to move. Thus the two quantities can have qualitatively different behavior.  We analyze the $S$-matrix model in the many-body limit (without recourse to mean field theory) in upcoming work~\cite{upcoming2}.

\section{Revisiting the literature} \label{sec.prev}

The superfluid weight has been computed from mean-field theory in a variety of multiband systems~\cite{Julku2016,Liang2017,Iskin2019,Wu2021,Iskin2019b,Peri2021,Chan2022,Herzog-Arbeitman2021,Kitamura2021,Peltonen2020}  including magic-angle twisted bilayer graphene~\cite{Julku2020,Xie2020,Hu2019} and flat band systems with disorder~\cite{Lau2022}. The impact of the terms arising from the derivatives of the order parameters should be examined on a case-by-case basis. For example, the results for the Lieb lattice presented in~[\onlinecite{Julku2016}] are mostly close to the correct result. Indeed, the hopping staggering $\delta$ used for the main results therein is very small and the orbital positions were picked so that $a=\frac{1}{2}$, which gives the correct results at $\delta=0$ even without including the derivatives of the order parameters. Results for larger values of $\delta$ are inaccurate. The results presented for the Mielke lattice with a flat band in~[\onlinecite{Iskin2019}] are accurate based on the same reasoning, but the results for other values of the tight-binding parameters may be affected by the ignored terms.

In Ref.~[\onlinecite{Chan2022}], the behavior of the order parameters was accurately taken into account, and the results agreed well with DMRG calculations. The superfluid weight was, however, compared with the quantum metric computed for a choice of the Fourier transformation which predicted $\pi D_s=0.6|U|$ at low interactions and for a half-filled flat band. The estimation we find using the correct choice, namely the minimal quantum metric, instead gives a slope of approximately $\pi D_s = 0.45|U|$, which is much closer to the mean-field and DMRG results of $\pi D_s\approx 0.40|U|$ obtained in~[\onlinecite{Chan2022}].

Expressions for the superfluid weight in terms of the quantum metric can be found in~[\onlinecite{Liang2017}] for models without flat bands. For instance, in the isolated band limit,
\begin{equation}
  [D_{s,{\rm geom}}]_{ij} = \frac{2}{V}\Delta^2\sum_{\vec{k}} \frac{{\rm
      tanh} (\beta E_{m,\vec{k}}/2)}{E_{m,\vec{k}}} {\rm Re}(\mathcal{B}_{ij}),
\end{equation}
where $m$ labels the isolated band, which does not need to be flat. In this case, the minimal quantum metric is not always relevant, but one should instead minimize the above integral for $i=j$.

The relationship between the superfluid weight and quantum metric has been used to derive various bounds for the superfluid weight~\cite{Peotta2015,Xie2020,Liang2017,Herzog-Arbeitman2021,Verma2021}. The lower bound given in~[\onlinecite{Peotta2015}] for time-reversal symmetric systems in terms of the spin Chern number is valid, as it is a lower bound for the quantum metric regardless of the choice of orbital positions. On the other hand, the bound proposed in~[\onlinecite{Liang2017}] related to the integral of the absolute value of the Berry curvature is only valid if one takes the lowest possible lower bound, as that quantity depends on the choice of orbital positions. This is also the case for the lower bound in terms of real space invariants proposed in~[\onlinecite{Herzog-Arbeitman2021}] for systems with obstructed Wannier orbitals or fragile topology. It is shown in the supplementary material of that work that the lower bound can be nonzero for arbitrary orbital positions. The correct choice of orbital positions is thus needed to define an orbital-independent bound. If the uniform pairing condition is satisfied, then space group symmetries can guarantee that the minimal quantum metric is obtained for orbitals at the high-symmetry positions.

The two-body problem in a flat band was shown in Ref.~[\onlinecite{Torma2018}] to give a finite effective mass for a pair, which means that already at the two-body level, interactions can lead to pair movement even when the single particle effective mass is infinite. The pair mass was found to be given by the ``local" (spatially dependent) version of the quantum metric -- which reassuringly is independent of orbital positions. However, approximations were then used to connect the pair mass to the usual quantum metric. Our many-body Cooper pair calculation in Section~\ref{sec.Jonah} now shows that the correct choice is the minimal quantum metric.  

Quantum geometry has been shown to be relevant also for Bose-Einstein condensation in flat bands~\cite{Julku2021,Julku20212}. The speed of sound and the excitation fraction were found to depend on generalized forms of the quantum metric, and the quantum distance between the flat band states, respectively. These quantities are invariant under the change of orbital positions. Under certain conditions, however, they were shown to reduce to the usual quantum metric and Hilbert-Smith quantum distance, and then (as well as in the superfluid density calculation in~[\onlinecite{Julku20212}]) one needs to pay attention to the choice of the correct basis.

Numerically exact methods such as quantum Monte Carlo do not require the same care as mean-field theory with the behavior of the order parameters, as the interaction Hamiltonian of the exact Hubbard model does not depend on the vector field explicitly. Generally, it is important to make sure that all variables that may depend on the vector potential are properly taken into account. 

In addition to the superfluid weight, the quantum metric has been related to the effective mass of two-body bounds states~\cite{Liang2018,Iskin2021,Iskin2022}, conductivity~\cite{Mitscherling2022}, the orbital magnetic susceptibility~\cite{Piechon2016,Gao2015}, the velocity of the Goldstone mode~\cite{Iskin2020}, and other physical phenomena~\cite{Abouelkomsan2022,Gao2019,Holder2020,Ahn2021,Mitscherling2020}. As shown here for the superfluid weight, whenever a connection is drawn between a physical quantity and the quantum metric, particular attention should be paid to the dependence of the quantum metric on the orbital positions. If the physical quantity should not depend on these, there may be an appropriate basis which is the only one where the quantum metric is relevant.

\subsection{Systems with broken time-reversal symmetry} \label{sec.no_trs}

Our result Eq.~\ref{eq.final} is valid for time-reversal symmetric systems. It can be straightforwardly generalized to be valid also for systems where TRS is broken (see Appendix~\ref{app.sfw}): the vector ${\rm d}_i\Delta$ will contain entries for the derivatives of the real parts of the order parameters and ${\rm d}\mu/{\rm d}q_i$ (when $\mu(\vec{q})\neq\mu(-\vec{q})$). Corresponding entries are added in the Hessian matrix $\partial_{\Delta,\mu}^2\Omega$. The addition of these terms should be considered carefully when connecting the superfluid weight to the quantum metric or other results obtained from $(1/V)\partial^2\Omega/\partial q_i\partial q_j\big|_{\vec{q}=\vec{0}}$. Indeed, in contrast to systems with TRS, there may not exist any set of orbital positions where $[D_s]_{ij} = (1/V)\partial^2\Omega/\partial q_i\partial q_j\big|_{\vec{q}=\vec{0}}$: the derivatives of the \textit{real parts} of the order parameters can be nonzero, and cannot be made zero by manipulating only the phases of the order parameters. If $[D_s]_{ij} = (1/V)\partial^2\Omega/\partial q_i\partial q_j\big|_{\vec{q}=\vec{0}}$ never holds, lower bounds derived for the quantum metric cannot generally be used for the superfluid weight. Furthermore, if ${\rm d}\mu/{\rm d}q_i\big|_{\vec{q}=\vec{0}}\neq 0$, the Hessian matrix $\partial_{\Delta,\mu}^2\Omega$ contains entries corresponding to the chemical potential and may not be positive semidefinite. In such a case, the partial derivative could even be smaller than the total derivative. 

When ${\rm d}\mu/{\rm d}q_i\big|_{\vec{q}=\vec{0}}=0$, the terms relating to $\mu$ can be ignored and the Hessian matrix $\partial_{\Delta}^2\Omega$ contains partial derivatives of $\Omega$ only with reference to the real and imaginary parts of the order parameters. In such a case, the inequality $\partial^2\Omega/\partial q_i^2\big|_{\vec{q}=\vec{0}}\geq {\rm d}^2F/{\rm d}q_i^2\big|_{\vec{q}=\vec{0}}$ holds. Furthermore, when the overall phase of the order parameters is fixed, $\partial_{\Delta}^2\Omega$ is invertible. Under these conditions, the superfluid weight is $[D_s]_{ij} = (1/V)\partial^2\Omega/\partial q_i\partial q_j\big|_{\vec{q}=\vec{0}}$ if and only if all the derivatives of the order parameters are zero at $\vec{q}=\vec{0}$. These derivatives are given by
\begin{equation}
    \frac{{\rm d}\Delta_{\alpha}}{{\rm d}q_i} = \frac{{\rm d}|\Delta_{\alpha}|}{{\rm d}q_i}e^{i\theta_{\alpha}} + i|\Delta_{\alpha}|e^{i\theta_{\alpha}}\frac{{\rm d}\theta_{\alpha}}{{\rm d}q_i}.
\end{equation}
Because changing the orbital positions only affects the phases $\theta_{\alpha}$, the derivatives can be set to zero with such a transformation only when ${\rm d}|\Delta_{\alpha}|/{\rm d}q_i=0\big|_{\vec{q}=\vec{0}}$. The equality $[D_s]_{ij} = (1/V)\partial^2\Omega/\partial q_i\partial q_j\big|_{\vec{q}=\vec{0}}$ can thus only hold in systems where ${\rm d}|\Delta_{\alpha}|/{\rm d}q_i\big|_{\vec{q}=\vec{0}}=0$ for all $\alpha,i$. In systems where ${\rm d}|\Delta_{\alpha}|/{\rm d}q_i\big|_{\vec{q}=\vec{0}} = 0$, results relating the superfluid weight to the quantum metric can be used, provided the diagonal components $(1/V)\partial^2\Omega/\partial q_i^2\big|_{\vec{q}=\vec{0}}$ are minimized.

\section{Conclusions} \label{sec.conc}

We have derived complete equations for the mean-field superfluid weight in multiband lattice models. These equations contain both the partial derivative of the grand potential, which gives a connection to quantum geometry, and terms that take into account the changes in the order parameter. The significance of the latter terms has been overlooked in the previous literature. We have shown that ignoring them can lead to quantitative as well as qualitative errors, where superconductivity can be predicted in systems where it is impossible. The use of the complete equations is thus crucial whenever studying multiband systems, such as moiré materials, as well as when searching for materials with particularly high critical temperatures.

Using our new equations, we have shown that the superfluid weight in isolated flat bands is proportional to the \textit{minimal} quantum metric, that is, the one with the smallest possible trace. A central discrepancy afflicting the current understanding of the connection between superconductivity and quantum geometry has been the following: the superfluid weight is manifestly independent on orbital positions, while the quantum metric, which has been shown to govern isolated flat band superconductivity, depends on them. Our finding that actually only the minimal quantum metric is relevant resolves this fundamental concern. Based on our results, bounds for the superfluid weight in terms of topological invariants in time-reversal symmetric systems~\cite{Peotta2015,Xie2020} are still valid, but other bounds which depend on the choice of orbital positions require more care. 

The conclusions based on the mean-field superfluid weight are corroborated by exact results derived for the Cooper pair mass. We generalized the uniform pairing condition in \Eq{eq:genUPC} to establish a minimal metric condition. When evaluated at the orbital positions satisfying the minimal metric condition, the Cooper pair mass is entirely determined by the quantum metric. Moreover, if the orbitals of the model are fixed by symmetries at high-symmetry points (maximal Wyckoff positions), then the minimal quantum metric is guaranteed to be obtained for these positions.

Importantly, our results show that in systems where TRS is broken, a relation between quantum geometry and superfluidity, and consequently topological bounds, does not exist in general. We identified sufficient conditions for having the connection to quantum geometry, namely that the derivatives of the order parameter and chemical potential with respect to $\vec{q}$ have to vanish at $\vec{q}=\vec{0}$. Whether these conditions are also necessary remains a topic of future research, as well as the possible relations of the conditions to the crystalline symmetries, as in the time-reversal symmetric, uniform pairing case.   

Furthermore, we have shown that the quantum geometry of the flat band is not sufficient to describe the superfluid weight in the non-isolated band limit: its behavior depends not only on the flat band properties but also on the nature of the band touching. In general, the geometric contribution is more prominent for linear band touchings than quadratic ones. Many flat band material candidates have band touchings~\cite{Regnault2021}. Remarkably, we have shown that an isolated flat band is \textit{not} necessary to achieve a high critical temperature, and that a band touching with dispersive bands can in fact be beneficial for superconductivity. This result is important for realizing the promise of high-temperature or even even room-temperature superconductivity from flat bands. Restricting to isolated flat bands would require materials and systems with a gap on the order tens of meV (the thermal energy). We have shown that this limitation is not necessary: in contrast, a band touching can enhance the critical temperature. This conclusion holds within the specific models considered by us, but is likely to be more general since the quantum metric of a flat band diverges when the gaps to the other bands are closed. By results from S-matrix analysis, we developed universal relations relating the pairing gaps on bipartite lattices, and argued that the pairing gap is enhanced for quadratic over linear band touchings, a result opposite to what we saw numerically for the superfluid weight. This is understood as density of states determining the former while also quantum geometry is important for the latter. Our results inspire further engineering of band touchings to optimize the critical temperature of superconductivity, and determine the dominance of  quantum geometry or the density of states.

\acknowledgments
We thank Aleksi Julku, Long Liang, Sebastiano Peotta, Grazia Salerno and Gabriel Topp for useful discussions. We acknowledge support by the Academy of Finland under project numbers 303351 and 327293. K-E.H. acknowledges support from the Magnus Ehrnrooth Foundation.   B.A.B. and A.C. were supported by the ONR Grant No. N00014-20-1-2303, DOE Grant No. DESC0016239, the Schmidt Fund for Innovative Research, Simons Investigator Grant No. 404513, the Packard Foundation, the Gordon and Betty Moore Foundation through Grant No. GBMF8685 towards the Princeton theory program, and a Guggenheim Fellowship from the John Simon Guggenheim Memorial Foundation. Further support was provided by the NSF-MRSEC Grant No. DMR-1420541 and DMR2011750, BSF Israel US foundation Grant No. 2018226, and the Princeton Global Network Funds.  B.A.B. acknowledges support from the Office of Naval Research grant No. N00014-20-1-2303 and from the European Research Council (ERC) under the European Union’s Horizon 2020 research and innovation programme (Grant agreement n° 101020833).  J.H-A. is supported by a Marshall Scholarship funded by the Marshall Aid Commemoration Commission.  A.C. is supported by a Moore Postdoctoral Fellowship from the Gordon and Betty Moore Foundation.

\phantom{a}\\


\appendix

\section{General equations for the superfluid weight} \label{app.sfw}

In this appendix, we derive the complete equations for the superfluid
weight without assuming time reversal symmetry. The Hamiltonian is invariant under a global change of phase of all order parameters, so we fix the overall phase by requiring that $\Delta_1$ is real and positive, with $\Delta_1$ a nonzero order parameter. We first apply the chain rule twice to the grand canonical potential to obtain
\begin{align}
  \frac{{\rm d}^2\Omega}{{\rm d}q_i{\rm d}q_j} &=
  \frac{{\rm d}}{{\rm d}q_j} \frac{\partial \Omega}{\partial q_i} +
  \frac{{\rm d}}{{\rm d}q_j} \left( \frac{\partial \Omega}{\partial
    \mu} \right) \frac{{\rm d}\mu}{{\rm d}q_i} + \frac{\partial
    \Omega}{\partial \mu} \frac{{\rm d}^2\mu}{{\rm d}q_i{\rm d}q_j}
  \nonumber \\
  &+\sum_{\alpha} \frac{{\rm d}}{{\rm d}q_j}\left(\frac{\partial
    \Omega}{\partial \Delta_{\alpha}^R}\right) \frac{{\rm
      d}\Delta_{\alpha}^R}{ {\rm d}q_i} + \sum_{\alpha} \frac{{\rm
      d}}{{\rm d} q_j} \left( \frac{\partial \Omega}{\partial
    \Delta_{\alpha}^I} \right) \frac{{\rm d}\Delta_{\alpha}^I}{{\rm
      d}q_i} \nonumber \\
  &+ \sum_{\alpha} \frac{\partial \Omega}{\partial \Delta_{\alpha}^R}
  \frac{{\rm d}^2\Delta_{\alpha}^R}{{\rm d}q_i{\rm d}q_j} +
  \sum_{\alpha} \frac{\partial \Omega}{\partial \Delta_{\alpha}^I}
  \frac{{\rm d}^2\Delta_{\alpha}^I}{{\rm d}q_i{\rm d}q_j}.
\end{align}
The particle number is fixed, meaning the second term on
the RHS of the first line is zero. The third term is canceled by the derivative
of $\mu N$ when taking the derivative of the free energy $F=\Omega
+\mu N$. Assuming that the order parameters solve the gap equation,
$\partial \Omega/\partial \Delta_{\alpha}=0$ for all
$\vec{q}$, the terms on the second and third lines all
vanish, and
\begin{align}
  \frac{{\rm d}^2F}{{\rm d}q_i{\rm d}q_j}  &=
  \frac{{\rm d}}{{\rm d}q_j} \frac{\partial \Omega}{\partial q_i} \\
  &= \frac{\partial^2\Omega}{\partial q_i\partial q_j} +
  \frac{\partial^2\Omega}{\partial \mu\partial q_i} \frac{{\rm
      d}\mu}{{\rm d}q_j} \nonumber \\
  &+ \sum_{\alpha} \left(
  \frac{\partial^2\Omega}{\partial \Delta_{\alpha}^R\partial q_i}
  \frac{{\rm d}\Delta_{\alpha}^R}{{\rm d}q_j}
  + \frac{\partial^2\Omega}{\partial \Delta_{\alpha}^I\partial q_i}
  \frac{{\rm d}\Delta_{\alpha}^I}{{\rm d}q_j} 
  \right) \bigg|_{\vec{q}=\vec{0}}.
\label{eq:totalsfw}
\end{align}
This equation can be written in a more compact form by using that
the particle number is kept fixed and $\partial
\Omega/\partial \Delta_{\alpha}=0$, implying that
\begin{equation}
  \frac{{\rm d}}{{\rm d}q_i} \frac{\partial \Omega}{\partial
    \Delta_{\alpha}^R} = \frac{{\rm d}}{{\rm d}q_i} \frac{\partial
    \Omega}{\partial 
    \Delta_{\alpha}^I} = \frac{{\rm d}}{{\rm d}q_i} \frac{\partial
    \Omega}{\partial \mu} = 0.  
\end{equation}
This system of equations can be written in matrix form as $(\partial_{\Delta,\mu}^2\Omega)\vec{f_i} = -\vec{b_i}$, where
\begin{widetext}
\begin{align}
  \partial_{\Delta,\mu}^2\Omega &= \begin{pmatrix}
    \frac{\partial^2 \Omega}{(\partial \Delta_1^R)^2} & \ldots &
    \frac{\partial^2 \Omega}{\partial \Delta_1^R\partial \Delta_n^R} &
    \frac{\partial^2\Omega}{ \partial \Delta_1^R\partial \Delta_2^I} &
    \ldots &
    \frac{\partial^2\Omega}{\partial\Delta_1^R\partial\Delta_n^I} &
    \frac{\partial^2\Omega}{\partial \Delta_1^R\partial\mu} \\
    \vdots & \ddots & \vdots & \vdots & \ddots & \vdots & \vdots \\
    \frac{\partial^2\Omega}{\partial \Delta_n^R\partial \Delta_1^R} &
    \ldots & \frac{\partial^2\Omega}{(\partial \Delta_n^R)^2} &
    \frac{\partial^2\Omega}{\partial \Delta_n^R\partial \Delta_2^I} &
    \ldots & \frac{\partial^2\Omega}{\partial
      \Delta_n^R\partial\Delta_n^I} & \frac{\partial^2\Omega}{\partial
      \Delta_{n}^R\partial \mu} \\
    \frac{\partial^2\Omega}{\partial \Delta_2^I\partial \Delta_1^R} &
    \ldots & \frac{\partial^2\Omega}{\partial \Delta_2^I\partial
      \Delta_n^R} & \frac{\partial^2\Omega}{(\partial \Delta_2^I)^2} &
    \ldots & \frac{\partial^2\Omega}{\partial \Delta_2^I\partial
      \Delta_n^I} & \frac{\partial^2\Omega}{\partial
      \Delta_2^I\partial \mu} \\
    \vdots & \ddots & \vdots & \vdots & \ddots & \vdots & \vdots \\
    \frac{\partial^2\Omega}{\partial \Delta_n^I\partial \Delta_1^R} &
    \ldots & \frac{\partial^2\Omega}{\partial \Delta_n^I\partial
      \Delta_n^R} & \frac{\partial^2\Omega}{\partial \Delta_n^I\partial
      \Delta_2^I} & \ldots & \frac{\partial^2\Omega}{(\partial
      \Delta_n^I)^2} & \frac{\partial^2\Omega}{\partial
      \Delta_n^I\partial \mu} \\
    \frac{\partial^2\Omega}{\partial \mu\partial \Delta_1^R} & \ldots &
    \frac{\partial^2\Omega}{\partial \mu\partial \Delta_n^R} &
    \frac{\partial^2\Omega}{\partial \mu\partial \Delta_2^I} & \ldots
    & \frac{\partial^2\Omega}{\partial \mu\partial \Delta_n^I} &
    \frac{\partial^2\Omega}{\partial \mu^2}
  \end{pmatrix}, \\
  \vec{f_i} &= \left( \frac{{\rm d}\Delta_1^R}{{\rm d}q_i}, \ldots,
  \frac{{\rm d}\Delta_n^R}{{\rm d}q_i}, \frac{{\rm d}\Delta_2^I}{{\rm
      d}q_i}, \ldots, \frac{{\rm d}\Delta_n^I}{{\rm d}q_i},
  \frac{{\rm d}\mu}{{\rm d}q_i} \right)^{\rm T} \\
  \vec{b_i} &= \left( \frac{\partial^2\Omega}{\partial q_i\partial
    \Delta_1^R}, \ldots, \frac{\partial^2\Omega}{\partial q_i\partial
    \Delta_n^R}, \frac{\partial^2\Omega}{\partial q_i\partial
    \Delta_2^I}, \ldots, \frac{\partial^2\Omega}{\partial q_i\partial
    \Delta_n^I}, \frac{\partial^2\Omega}{\partial q_i\partial \mu}
  \right)^{\rm T}. 
\end{align}
\end{widetext}
The order parameter $\Delta_1^I$ is absent as we have set the global phase of all order parameters by forcing $\Delta_1$ real and positive.
With these definitions, the total superfluid weight in \Eq{eq:totalsfw} can be written as
\begin{equation}
  V[D_s]_{ij} = \frac{\partial^2\Omega}{\partial q_i\partial q_j}
  \bigg|_{\vec{q}=\vec{0}} - \vec{f_i}^{\rm T} (\partial_{\Delta,\mu}^2\Omega) \vec{f_i}\big|_{\vec{q}=\vec{0}}.
\end{equation}
The derivatives of the order parameters and chemical potential can be
found by solving the state at nonzero $\vec{q}$ or from the system of
equations $(\partial_{\Delta,\mu}^2\Omega)\vec{f_i} = -\vec{b_i}$ if the matrix $\partial_{\Delta,\mu}^2\Omega$ is invertible. If we had not fixed the overall phase of the order parameters, $\partial_{\Delta,\mu}^2\Omega$ would be singular. However, removing the line and column involving derivatives with reference to $\Delta_1^I$ from the Hessian matrix as we have done in the definition of $\partial_{\Delta,\mu}^2\Omega$ generally makes $\partial_{\Delta,\mu}^2\Omega$ non-singular.

When the derivatives of the order parameters are purely imaginary, for example in systems with TRS, the additional terms $- \vec{f_i}^{\rm T} (\partial_{\Delta,\mu}^2\Omega) \vec{f_i}\big|_{\vec{q}=\vec{0}}$ appear only in multiband models. However, if the real part of the order parameters has a nonzero derivative, $[D_s]_{ij} = (1/V)\partial \Omega/{\partial}q_i{\partial}q_j\big|_{\vec{q}=\vec{0}}$ can be inaccurate even in single-band models, as the derivative cannot be made zero by changing the phase of the order parameter.

Here we have used only the order parameters as mean-field parameters. If one included more parameters, for example Hartree terms, the $\vec{q}$ dependence of those parameters should be appropriately taken into account as well. 

We note here that when ${\rm d}\mu/{\rm d}q_i\big|_{\vec{q}=\vec{0}}\neq 0$, 
the derivatives of the chemical potential may
contribute to the superfluid weight. Both the definition $[D_s]_{ij} =
(1/V){\rm d}^2\Omega/{\rm d}q_i{\rm d}q_j|_{\mu,\vec{q}=\vec{0}}$, where
$\mu$ is fixed, and $[D_s]_{ij} = (1/V){\rm d}^2F/{\rm d}q_i{\rm d}q_j\big|_{N,\vec{q}=\vec{0}}$ used above have been used in literature, but it is unclear whether they always yield the same result at the mean-field level. This ambiguity is related to the
non-conservation of the particle number by the BCS Hamiltonian, which makes
the introduction of the chemical potential more subtle at the
mean-field level than in the exact Hubbard Hamiltonian. If $\mu$ is
thought of as a Lagrange multiplier that should be solved to keep the
average particle number constant, its dependence on $\vec{q}$ should
be included.

\section{Impact of orbital positions on the order parameters} \label{app.orb_pos}

With our convention of Fourier transformation
(Eq.~\eqref{eq.fourier}), the intra-unit cell orbital positions
$\vec{\delta_{\alpha}}$ appear in the Fourier transformed kinetic
Hamiltonians, $[H_{\vec{k}}^{\sigma}]_{\alpha\beta} = -\sum_i
t_{i\alpha,0\beta}^{\sigma} e^{-i\vec{k}\cdot
  (\vec{R_i}+\vec{\delta_{\alpha}}-\vec{\delta_{\beta}})}$. Let us
denote by $\widetilde{H}^{\sigma}_{\vec{k}}$ and
$H^{\sigma}_{\vec{k}}$ the kinetic Hamiltonians with intra-unit cell 
positions $\{\widetilde{\vec{\delta_{\alpha}}}\}$ and $\{\vec{\delta_{\alpha}}\}$,
respectively. The two Hamiltonians are related by
\begin{align}
  [\widetilde{H}^{\sigma}_{\vec{k}}]_{\alpha\beta} &= -
  e^{-i\vec{k}\cdot(\widetilde{\vec{\delta_{\alpha}}} -
    \widetilde{\vec{\delta_{\beta}}})} \sum_i t_{i\alpha, 0\beta}^{\sigma}
  e^{-i\vec{k} \cdot \vec{R_i}} \nonumber \\
  &= e^{-i\vec{k}\cdot(
    \widetilde{\vec{\delta_{\alpha}}} - \vec{\delta_{\alpha}} 
    -\widetilde{\vec{\delta_{\beta}}} + \vec{\delta_{\beta}})}
  [H^{\sigma}_{\vec{k}}]_{\alpha\beta}. 
\end{align}
This can be rewritten in matrix form as 
$\widetilde{H}^{\sigma}_{\vec{k}} =
\crea{V_{\vec{k}}}H^{\sigma}_{\vec{k}}\ani{V_{\vec{k}}}$, where
$V_{\vec{k}} = {\rm
  diag}(e^{i\vec{k}\cdot(\widetilde{\vec{\delta_{1}}} -
  \vec{\delta_{1}})}, \ldots, e^{i\vec{k}\cdot(\widetilde{\vec{\delta_{n}}} -
  \vec{\delta_{n}})})$.

To show how the orbital positions impact the order parameters, let us
consider the corresponding Bogoliubov-de-Gennes (BdG) Hamiltonians. By
performing a unitary transformation
$\ani{U}\widetilde{H}_{BdG}(\vec{k})\crea{U}$ with $U = {\rm diag}
(V_{\vec{q}+\vec{k}}, V_{\vec{q}-\vec{k}}^{\dag})$,  
$\widetilde{H}_{BdG}(\vec{k})$ becomes 
\begin{equation}
  \ani{U}\widetilde{H}_{BdG,\vec{k}}\crea{U} = \begin{pmatrix}
    H_{\vec{q}+\vec{k}}^{\up} - \mu\vec{1} & V_{\vec{q}+\vec{k}}
    \vec{\Delta} V_{\vec{q}-\vec{k}} \\
    V_{\vec{q}-\vec{k}}^{\dag}\vec{\Delta}^{\dag}V_{\vec{q}+\vec{k}}^{\dag} &
    -\left( H^{\dn}_{\vec{q}-\vec{k}} \right)^* + \mu\vec{1}
  \end{pmatrix}.
\end{equation}
Assuming $\vec{\Delta}$ is diagonal, it commutes with $V$, and
$V_{\vec{q}+\vec{k}} \vec{\Delta} V_{\vec{q}-\vec{k}} = {\rm diag}
(\Delta_1 e^{2i\vec{q}\cdot(\widetilde{\vec{\delta_1}} -
  \vec{\delta_1})},\ldots,\Delta_n
e^{2i\vec{q}\cdot(\widetilde{\vec{\delta_n}} - 
  \vec{\delta_n})} )$. Thus $\widetilde{H}_{BdG}(\vec{k})$ with order
parameters $(\Delta_1,\ldots,\Delta_n)$ has the same eigenvalues as
$H_{BdG}(\vec{k})$ with order parameters $(\Delta_1
e^{2i\vec{q}\cdot(\widetilde{\vec{\delta_1}} -   \vec{\delta_1})},\ldots,\Delta_n
e^{2i\vec{q}\cdot(\widetilde{\vec{\delta_n}} -  
  \vec{\delta_n})} )$. Since the grand canonical potential depends
only on the eigenvalues of the BdG Hamiltonian and the absolute value
of the order parameters, the thermodynamic potentials are related by 
\begin{equation}
  \widetilde{\Omega}(\vec{q},\mu,\Delta_{\alpha}) =
  \Omega(\vec{q},\mu,\Delta_{\alpha}e^{2i\vec{q}\cdot(\widetilde{\vec{\delta_{\alpha}}} 
    -   
  \vec{\delta_{\alpha}})}). \label{eq.relation}
\end{equation}
The thermodynamic potential at the order parameters that solve the gap
equation will always be the same for a given $\vec{q}$ and $\mu$
regardless of the intra-unit cell positions. However, the order
parameters that minimize the thermodynamic potential will have complex
phases that depend on the intra-unit cell positions. These phases can
be sublattice-dependent, and in the multiband case, they cannot in
general be removed by a change in the overall phase of the order
parameters.

\section{Positions for which the superfluid weight is related to the
  quantum metric} \label{app.unique}

The superfluid weight in a system with TRS is given by the simple equation $[D_s]_{ij} =
(1/V)\partial^2\Omega/\partial q_i\partial q_j\big|_{\vec{q}=\vec{0}}$ when $-({\rm
  d}_i\Delta^{I})^{T}\partial_{\Delta^I}^2\Omega({\rm d}_j\Delta^{I})\big|_{\vec{q}=\vec{0}}  = 0$ for all $i,j$. When $\partial_{\Delta^I}^2\Omega$ is invertible, this holds if and only if 
${\rm d}_i{\Delta^I}=0$. This is the case when the overall phase of the order parameters is fixed.

The derivatives of the order parameters in a system with TRS are given
by
\begin{equation}
  \frac{{\rm d}\Delta_{\alpha}}{{\rm d}q_i}\bigg|_{\vec{q}=\vec{0}} = i\frac{{\rm
    d}\Delta_{\alpha}^I}{{\rm d}q_i}\bigg|_{\vec{q}=\vec{0}} 
= i\Delta_{\alpha}\frac{{\rm d}\theta_{\alpha}}{{\rm d}q_i} \bigg|_{\vec{q}=\vec{0}}.
\label{eq:thetapositions}
\end{equation}
As shown in Appendix~\ref{app.orb_pos}, if the solutions to the gap
equation in a system with orbital positions
$\{\vec{\delta_{\alpha}}\}$ are
$\Delta_{\alpha}=|\Delta_{\alpha}|e^{i\theta_{\alpha}}$, the solutions
with another choice of positions $\{\vec{\delta_{\alpha}}'\}$ are
$\Delta_{\alpha}' = |\Delta_{\alpha}|e^{i\theta_{\alpha}'}$, where
$\theta_{\alpha}' = \theta_{\alpha} - 2\vec{q}\cdot
(\vec{\delta_{\alpha}}' - \vec{\delta_{\alpha}})$. The derivatives of
the order parameters are thus related by
\begin{equation}
  \frac{{\rm d}\Delta_{\alpha}^{I}}{{\rm d}q_i}\bigg|_{\vec{q}=\vec{0}}
  = \Delta_{\alpha}\frac{{\rm d}\theta_{\alpha}}{{\rm d}
    q_i}\bigg|_{\vec{q}=\vec{0}} = \Delta_{\alpha}\frac{{\rm
      d}\theta_{\alpha}'}{{\rm d}q_i}\bigg|_{\vec{q}=\vec{0}} +
  2\Delta_{\alpha}[\vec{\delta_{\alpha}}'-\vec{\delta_{\alpha}}]_i. \label{eq.app_rel} 
\end{equation}
The positions $\vec{\delta_{\alpha}}$ for which ${\rm
  d}_i\vec{\Delta}^I=0$ can be solved directly from this equation once
the derivative is known for some positions
$\{\vec{\delta_{\alpha}^0}\}$. When $\partial_{\Delta^I}^2\Omega$ is invertible, the
derivatives of the order parameters are uniquely defined, and the
above equation gives a unique position $[\vec{\delta_{\alpha}}]_i=[\vec{\delta_{\alpha}^{0}}]_i + ({\rm d}\theta_{\alpha}^0/{\rm d}q_i)/2\big|_{\vec{q}=\vec{0}}$ for all sublattices where
$\Delta_{\alpha}\neq 0$. 

The initial choice of orbital positions $\{\vec{\delta}_{\alpha}^{0}\}$ is arbitrary, and we can verify that the solution $\{\vec{\delta_{\alpha}}\}$ where ${\rm d}\Delta^I=0$ remains the same with a different choice. If we pick
another initial set of positions $\{\vec{\delta}_{\alpha}^{1}\}$, the
positions for which ${\rm d}\Delta_{\alpha}/{\rm
  d}q_i\big|_{\vec{q}=\vec{0}} = 0$ are
\begin{align}
  [\vec{\delta_{\alpha}}]_i &= [\vec{\delta_{\alpha}^1}]_i +
  \frac{1}{2}\frac{{\rm d}\theta_{\alpha}^{1}}{{\rm
      d}q_i}\bigg|_{\vec{q}=\vec{0}}
  = [\vec{\delta_{\alpha}^1}]_i -
  [\vec{\delta_{\alpha}^1}-\vec{\delta_{\alpha}^0}]_i +
  \frac{1}{2}\frac{{\rm d}\theta_{\alpha}^0}{{\rm
                              d}q_i}\bigg|_{\vec{q}=\vec{0}} \nonumber \\
  &= [\vec{\delta_{\alpha}^0}]_i +
  \frac{1}{2}\frac{{\rm d}\theta_{\alpha}^0}{{\rm 
      d}q_i}\bigg|_{\vec{q}=\vec{0}},
\end{align}
for any sublattice $\alpha$ where $\Delta_{\alpha}\neq 0$. We used
Eq.~\eqref{eq.app_rel} in the second equality. The positions $\{\vec{\delta_{\alpha}}\}$ are thus the same
for any choice of initial orbital positions.

If we had not fixed the overall phase of the parameters at nonzero $\vec{q}$, the vector ${\rm d}_i\Delta^{I}$ and the Hessian matrix $\partial_{\Delta^I}^2\Omega$ would read
\begin{align}
  {\rm d}_i\Delta^I &= \left( \frac{{\rm d}\Delta_1^I}{{\rm d}q_i},
  \ldots, \frac{{\rm d}\Delta_{n}^I}{{\rm d} q_i}
  \right)^{\rm T},\\
  \partial_{\Delta^I}^2\Omega &= \begin{pmatrix}
    \frac{\partial^2\Omega}{\partial \Delta_{1}^I\partial
      \Delta_{1}^I} & \ldots & \frac{\partial^2\Omega}{\partial
      \Delta_{1}^I\partial \Delta_{n}^I} \\
    \vdots & \ddots & \vdots \\
    \frac{\partial^2\Omega}{\partial \Delta_{n}^I\partial \Delta_1^I}
    & \ldots & \frac{\partial^2\Omega}{\partial \Delta_n^I\partial
      \Delta_n^I}
    \end{pmatrix}.
\end{align}
These have the same form as in the main text, but with the addition of
the terms related to $\Delta_1$. The full Hessian matrix is not
invertible, but has an eigenvector $\vec{v}=(\Delta_1,\ldots,\Delta_n)^T$ with
a zero eigenvalue, which reflects the freedom in the phase of the order parameters~\cite{Chan2022}. In this 
case, $-({\rm
  d}_i\Delta^{I})^{T}\partial_{\Delta^I}^2\Omega({\rm d}_j\Delta^{I})=0$ if
and only if ${\rm d}_i\Delta^{I} = C_i\vec{v}$, where $C_i$ is a
real number. Then from \Eq{eq:thetapositions}, the positions for which $[D_s]_{ij} =
(1/V)\partial^2\Omega/\partial q_i\partial
q_j\big|_{\vec{q}=\vec{0}}$ are given by
\begin{equation}
  [\vec{\delta_{\alpha}}]_i = [\vec{\delta_{\alpha}^{0}}]_i +
  \frac{1}{2} \frac{{\rm d}\theta_{\alpha}^0}{{\rm
      d}q_i}\bigg|_{\vec{q}=\vec{0}} + C_i,
\end{equation}
in sublattices where $\Delta_{\alpha}\neq 0$. Like before,
$\{\vec{\delta_{\alpha}^0}\}$ are arbitrary orbital positions. If the
overall phase of the order parameters is not fixed, the positions for
which $[D_s]_{ij} =
(1/V)\partial^2\Omega/\partial q_i\partial q_j\big|_{\vec{q}=\vec{0}}$ are thus uniquely defined up to an
overall translation by $C_i$.

\section{Superfluid weight from linear response theory}~\label{app.lin}

When computing the mean-field superfluid weight from the current
response as in~[\onlinecite{Liang2017}], we get the same result as
from $D_s = (1/V)\partial^2\Omega/\partial q_i\partial q_j\big|_{\vec{q}=\vec{0}}$. This is expected, as the dependence of the order parameters on the vector
field is ignored. Here we compute the superfluid weight from linear
response theory by taking this dependence into account, and obtain an
expression that is equivalent with $[D_s]_{ij} = (1/V){\rm d}^2F/{\rm
  d}q_i{\rm d}q_j\big|_{\vec{q}=\vec{0}}$ when ${\rm d}\mu/{\rm d}q_i\big|_{\vec{q}=\vec{0}}=0$. 

Let us start from the mean-field Hamiltonian
\begin{align}
  H_{\rm MF} &= H_{\rm kin} + H_{\rm int}, \\
  H_{\rm int} &= \sum_{i\alpha} \Delta_{i\alpha}
  \crea{c_{i\alpha\up}}\crea{c_{i\alpha\dn}} +
  \Delta_{i\alpha}^*\ani{c_{i\alpha\dn}} \ani{c_{i\alpha\up}} -
  \frac{|\Delta_{i\alpha}|^2}{U},
\end{align}
where $\Delta_{i\alpha} =
U\ave{\ani{c_{i\alpha\dn}}\ani{c_{i\alpha\up}}}$. The vector field is
introduced using the standard Peierls substitution in the kinetic term, so that
$t_{i\alpha,j\beta}^{\sigma}$ is rewritten as
$t_{i\alpha,j\beta}^{\sigma}(\vec{A}) = t_{i\alpha,j\beta}^{\sigma}
{\rm exp}\left( -i\int_{\vec{r}_{i\alpha}}^{\vec{r}_{j\beta}}
  \vec{A}\cdot {\rm d}\vec{r}\right)$. We assume that $\vec{A}$ varies slowly in
space and time. Then the hopping terms can be approximated by
$t_{i\alpha,j\beta}^{\sigma}(\vec{A}) =
t_{i\alpha,j\beta}^{\sigma}e^{-i\vec{A}(\vec{r}^{\rm
    CM}_{i\alpha,j\beta},t)\cdot \vec{r}^{\rm rel}_{i\alpha,j\beta}}$,
where $\vec{r}^{\rm rel}_{i\alpha,j\beta} 
= \vec{r}_{i\alpha}-\vec{r}_{j\beta}$ and $r_{i\alpha,j\beta}^{\rm CM} =
(\vec{r}_{i\alpha}+\vec{r}_{j\beta})/2$.  The total current density
induced by $\vec{A}$ is $j_{\mu}(\vec{r},t)
= -\delta H(\vec{A})/\delta A_{\mu}(\vec{r},t)$, where $\delta/\delta
A_{\mu}$ is 
the functional derivative with reference to $A_{\mu}$. 

We first expand the kinetic term up to second order in $\vec{A}$ around $\vec{A}=\vec{0}$ to obtain the functional derivative up to first order:
\begin{equation}
  \frac{\delta H_{\rm kin}(\vec{A})}{\delta A_{\mu}(\vec{r},t)} =
  \sum_{\substack{
      i\alpha,j\beta\\\vec{r}_{i\alpha,j\beta}^{\rm CM} = \vec{r}}
    }
  T_{\mu\nu}(i\alpha,j\beta)A_{\nu}(\vec{r},t) + j_{\mu}^{\rm
    p}(i\alpha,j\beta). \label{eq.cur_kin_term}
\end{equation}
Repeated indices are summed over. The operators
$T_{\mu\nu}(i\alpha,j\beta)A_{\nu} = -\sum_{\sigma}
t_{i\alpha,j\beta}^{\sigma} [r_{i\alpha,j\beta}^{\rm rel}]_{\mu}
[r_{i\alpha,j\beta}^{\rm
  rel}]_{\nu}\crea{c_{i\alpha\sigma}}\ani{c_{j\beta\sigma}}A_{\nu}$
and $j_{\mu}^{\rm p}(i\alpha,j\beta) = -i\sum_{\sigma}
t_{i\alpha,j\beta}^{\sigma} [r_{i\alpha,j\beta}^{\rm
  rel}]_{\mu}\crea{c_{i\alpha\sigma}}\ani{c_{j\beta\sigma}}$ are the
diamagnetic and paramagnetic current operators, respectively. 

The functional derivative of the mean-field interaction Hamiltonian is
\begin{equation}
    \frac{\delta H_{\rm int}}{\delta A_{\mu}} = \sum_{i\alpha}  \frac{\delta \Delta_{i\alpha}}{\delta A_{\mu}} \crea{c_{i\alpha\up}}\crea{c_{i\alpha\dn}} + {\rm H.c.} - \left( \frac{\delta \Delta_{i\alpha}}{\delta A_{\mu}} \frac{\Delta_{i\alpha}^*}{U}+{\rm H.c.}\right). \label{eq.int_fun_der}
\end{equation}
Using the linear response approximation $\Delta_{i\alpha}(\vec{A})\approx \Delta_{i\alpha}(\vec{A}=\vec{0}) + \delta \Delta_{i\alpha}/\delta A_{\nu}\big|_{\vec{A}=\vec{0}}A_{\nu}$, Eq.~\eqref{eq.int_fun_der} becomes
\begin{align}
    \frac{\delta H_{\rm int}}{\delta A_{\mu}} =& \sum_{i\alpha}  \frac{\delta \Delta_{i\alpha}}{\delta A_{\mu}} \crea{c_{i\alpha\up}}\crea{c_{i\alpha\dn}}\bigg|_{\vec{A}=\vec{0}} + {\rm H.c.} \nonumber \\
    -& \frac{1}{U}\sum_{i\alpha}\left(\Delta_{i\alpha}^*\frac{\delta \Delta_{i\alpha}}{\delta A_{\mu}}\bigg|_{\vec{A}=\vec{0}} + {\rm H.c} \right) \nonumber \\
    -& \frac{1}{U}\sum_{i\alpha} \left(\frac{\delta \Delta_{i\alpha}}{\delta A_{\mu}}\frac{\delta \Delta_{i\alpha}^*}{\delta A_{\nu}}\bigg|_{\vec{A}=\vec{0}} + {\rm H.c.}\right)A_{\nu}. 
    \label{eq.cur_int_term}
\end{align} 
By combining equations~~\eqref{eq.cur_kin_term} and \eqref{eq.cur_int_term}, we obtain the total current density operator
\begin{widetext}
\begin{align}
  \ave{j_{\mu}(\vec{r},t)} &=
  - \sum_{\substack{i\alpha,j\beta:\\\vec{r}_{i\alpha,j\beta}^{\rm CM}
      = \vec{r}}} \left[ \ave{\widetilde{T}_{\mu\nu}(i\alpha,j\beta)}
  A_{\nu}(\vec{r}_{i\alpha,j\beta}^{\rm CM}, t) + \ave{\widetilde{j_{\mu}^{\rm
        p}}(i\alpha,j\beta)} \right],\\
  \widetilde{T}_{\mu\nu}(i\alpha,j\beta) &= T_{\mu\nu}(i\alpha,j\beta)
  - \frac{1}{U}\left(\frac{\delta \Delta_{i\alpha}}{\delta
    A_{\mu}(\vec{r}_{i\alpha},t)} \frac{\delta
    \Delta^*_{i\alpha}}{\delta A_{\nu}(\vec{r}_{i\alpha},t)}\bigg|_{\vec{A}=\vec{0}} + {\rm
      H.c.} \right) \delta_{i\alpha,j\beta} \\
  \widetilde{j_{\mu}^{\rm p}}(i\alpha,j\beta) &= j_{\mu}^{\rm
    p}(i\alpha,j\beta) 
  +\left( \frac{\delta\Delta_{i\alpha}}{\delta
    A_{\mu}(\vec{r}_{i\alpha},t)} \bigg|_{\vec{A}=\vec{0}}
  \crea{c_{i\alpha\up}}\crea{c_{i\alpha\dn}} -
  \frac{1}{U}\frac{\delta 
    \Delta_{i\alpha}}{\delta A_{\mu}(\vec{r}_{i\alpha},t)}
  \Delta_{i\alpha}^*\bigg|_{\vec{A}=\vec{0}} + {\rm
    H.c.} \right) \delta_{i\alpha,j\beta}.
    \end{align}
\end{widetext}

As $\vec{A}$ varies slowly in both space and time, we can assume the
induced current has the same spatial and temporal dependence as
$\vec{A}$, so that
\begin{equation}
  \ave{j_{\mu}(\vec{q},\omega)} =
  -K_{\mu\nu}(\vec{q},\omega)A_{\nu}(\vec{q},\omega),
\end{equation}
where $K_{\mu\nu}$ is the current-current response function.
The Fourier transformed total current density reads $
\ave{j_{\mu}(\vec{q},t)} = (1/V)\sum_{\vec{r}} \ave{j_{\mu}(\vec{r},t)}
e^{-i\vec{q}\cdot \vec{r}}$. Assuming the order parameter is uniform in each sublattice (i.e., may depend on the orbital $\alpha$, but for a given orbital is the same at each unit cell $i$), we obtain
\begin{align}
  \ave{j_{\mu}(\vec{q},t)} &=
  -\ave{\widetilde{T}_{\mu\nu}}A_{\nu}(\vec{q},t) - 
  \ave{\tilde{j_{\mu}^{\rm p}}(\vec{q})},\\
  \widetilde{T}_{\mu\nu} &=\frac{1}{V} \sum_{\vec{k},\sigma}\sum_{\alpha\beta}
  [\partial_\mu\partial_\nu H_{\sigma}(\vec{k}')|_{\vec{k'=\vec{k}}}]_{\alpha\beta}
  \crea{c_{\vec{k}\alpha\sigma}} \ani{c_{\vec{k}\beta\sigma}}
  \nonumber \\
  &-\frac{1}{U}\frac{1}{V_c}\sum_{\alpha} \left( \frac{\delta
    \Delta_{\alpha}}{\delta 
    A_{\mu}} \frac{\delta \Delta_{\alpha}^*}{\delta A_{\nu}}
  + {\rm 
    H.c. } \right), \\
  \widetilde{j_{\mu}^{\rm p}}(\vec{q}) &=
  \frac{1}{V} \sum_{\vec{k},\sigma}\sum_{\alpha\beta} 
  [\partial_{\mu} H_{\sigma}(\vec{k}')|_{\vec{k}'=\vec{k}+\vec{q}/2}]_{\alpha\beta}
  \crea{c_{\vec{k}\alpha\sigma}} \ani{c_{\vec{k}+\vec{q}\beta\sigma}}
  \nonumber \\
  &+ \frac{1}{V}\sum_{\vec{k}\alpha} \frac{\delta
    \Delta_{\alpha}}{\delta A_{\mu}} 
  \crea{c_{\vec{k}-\vec{q} \alpha \up}} \crea{c_{-\vec{k} \alpha\dn}} 
  + \frac{\delta \Delta_{\alpha}^*}{\delta A_{\mu}}
  \ani{c_{-\vec{k} \alpha \dn}} \ani{c_{\vec{k}+\vec{q} \alpha\up}}
  \nonumber \\
  &- \frac{1}{V}\sum_{\vec{r}_{i\alpha}}\frac{1}{U}\left(\frac{\delta
      \Delta_{\alpha}^*}{\delta A_{\mu}} \Delta_{\alpha}(\vec{0}) +
  {\rm H.c.} 
  \right)e^{-i\vec{q}\cdot\vec{r}_{i\alpha}},  \label{eq.gen_para}
\end{align}
where $\delta \Delta_{\alpha}/\delta A_{\mu}=\delta
\Delta_{i\alpha}/\delta A_{\mu}(\vec{r}_{i\alpha},t)\big|_{\vec{A}=\vec{0}}$ and $V_c$ is the
volume of a unit cell, $V_c = V/N_c$. 

In linear response theory, the paramagnetic part can be computed
using the Kubo formula
\begin{widetext}
\begin{equation}
  \ave{\widetilde{j_{\mu}^{\rm p}}(\vec{q},\omega)} = 
-i V \sum_{\nu} \int_{0}^{\infty} {\rm d}t e^{i\omega t} \ave{[
  \widetilde{j_{\mu}^{\rm p}}(\vec{q},t), \widetilde{j_{\nu}^{\rm p}}
  (-\vec{q}, 0)
]} A_{\nu}(\vec{q},\omega).\label{eq.para_kubo}
\end{equation}
\end{widetext}

We will compute the current-current response function $K_{\mu\nu}$ in imaginary time using the Matsubara formalism. To compute the contribution from the paramagnetic current, we define
\begin{equation}
  \Pi_{\mu\nu}(\vec{q},\tau) = V^2\ave{T[\widetilde{j_{\mu}^{\rm
        p}}(\vec{q},\tau) \widetilde{j_{\nu}^{\rm p}}(-\vec{q},0)]}, 
\end{equation}
where $T$ is the imaginary time ordering operator.

In the computation of $\Pi_{\mu\nu}$, it will be useful to define the
following block matrices:
\begin{align}
  \widetilde{H}(\vec{k}) &= \begin{pmatrix}
    H_{\up}(\vec{k}) & \vec{0} \\ \vec{0} & -H_{\dn}^*(-\vec{k})
  \end{pmatrix}, \\
  G^{\alpha\beta}(\vec{k}) &= -\begin{pmatrix}
    \ave{T[\ani{c_{\vec{k}\alpha\up}}(\tau)
      \crea{c_{\vec{k}\beta\up}}]}
    & \ave{T[\ani{c_{\vec{k}\alpha\up}}(\tau)
      \ani{c_{-\vec{k}\beta\dn}}]} \\
    \ave{T[\crea{c_{-\vec{k}\alpha\dn}}(\tau)
      \crea{c_{\vec{k}\beta\up}}]} &
    \ave{T[\crea{c_{-\vec{k}\alpha\dn}}(\tau)
      \ani{c_{-\vec{k}\beta\dn}}]}
  \end{pmatrix}, \\
  \delta_{\nu} \Delta &= \begin{pmatrix}
    \vec{0} & \frac{\delta \vec{\Delta}}{\delta A_{\nu}} \\
    \frac{\delta \vec{\Delta}^*}{\delta A_{\nu}} & \vec{0} 
  \end{pmatrix}, \\
  \frac{\delta \vec{\Delta}}{\delta A_{\nu}} &= {\rm diag}\left(
    \frac{\delta \Delta_1}{\delta A_{\nu}}, \ldots, \frac{\delta
    \Delta_n}{\delta A_{\nu} } \right).
\end{align}
We use the following indexing convention: $A_{ij}$ designates
the block $(i,j)$, and $A_{ij}^{\alpha\beta}$ designates
the component $(\alpha,\beta)$ in said block. For example,
$G(\tau,\vec{k})^{\alpha\beta}_{01} =
-\ave{T[\ani{c_{\vec{k}\alpha\up}}\ani{c_{-\vec{k}\beta\dn}}]}$. For
readability, we will use the notation $\partial_{\mu} A|_{\vec{k}} =
\partial A(\vec{k}')/\partial k_{\mu}'|_{\vec{k}'=\vec{k}}$.

If we do not take the dependence of order parameters into account, the
only terms in $\Pi_{\mu\nu}(\vec{q},\tau)$ are
\begin{align}
  \sum_{\vec{k}\vec{k}'\sigma\sigma'\alpha\beta\gamma\delta}
  &[\partial_{\mu}
  H_{\sigma}|_{\vec{k}+\vec{q}/2}]^{\alpha\beta} 
  [\partial_{\nu}
  H_{\sigma'}|_{\vec{k}'-\vec{q}/2}]^{\gamma\delta}
  \nonumber \\ 
  &\ave{T[\crea{c_{\vec{k}\alpha\sigma}}(\tau)\ani{c_{\vec{k}+
        \vec{q}\beta\sigma}}(\tau)  
    \crea{c_{\vec{k}'\gamma\sigma'}}
    \ani{c_{\vec{k}'-\vec{q}\delta\sigma'}}]}. 
\end{align}
These can be expressed as $\Pi^{(0)}_{\mu\nu} = -\sum_{\vec{k}} {\rm
  Tr}[G(-\tau,\vec{k}) 
\partial_{\mu} \widetilde{H}|_{\vec{k}+\vec{q}/2}\gamma^z
G(\tau,\vec{k}+\vec{q}) \partial_{\nu}\widetilde{H}|_{\vec{k}+\vec{q}/2}\gamma^z]$.

For the new terms related to the derivatives of the order parameters, 
let us start from those where the prefactor
involves one derivative of $\Delta_{\alpha}$ or
$\Delta_{\alpha}^*$. We will show detailed steps for 
\begin{align}
  \sum_{\vec{k}\vec{k}'\sigma\alpha\beta\gamma} &\frac{\delta
    \Delta_{\alpha}}{\delta A_{\mu}} [\partial_{\nu} 
    H_{\sigma}|_{\vec{k}'-\vec{q}/2}]^{\beta\gamma} \nonumber \\
  &\ave{
    T[\crea{c_{\vec{k}-\vec{q}\alpha\up}}(\tau)
    \crea{c_{-\vec{k}\alpha\dn}}(\tau)
    \crea{c_{\vec{k}'\beta\sigma}}\ani{c_{\vec{k}'-\vec{q}\gamma\sigma}}] \label{eq.ex1}
  }.
\end{align}
Taking only one-loop graphs and ignoring disconnected ones, the four
point correlator becomes
\begin{widetext}
\begin{align}
  &\ave{T[\crea{c_{\vec{k}-\vec{q}\alpha\up}}(\tau)
  \crea{c_{-\vec{k}\alpha\dn}}(\tau)
  \crea{c_{\vec{k}'\beta\sigma}}\ani{c_{\vec{k}'-\vec{q}\gamma\sigma}}]} \nonumber\\
  &=
  -\ave{T[\crea{c_{\vec{k}-\vec{q}\alpha\up}}(\tau)
    \crea{c_{\vec{k}'\beta\sigma}}]}
  \ave{T[\crea{c_{-\vec{k}\alpha\dn}}(\tau)
    \ani{c_{\vec{k}'-\vec{q}\gamma\sigma}}
  ]} +
  \ave{T[\crea{c_{\vec{k}-\vec{q}\alpha\up}}(\tau)
    \ani{c_{\vec{k}'-\vec{q}\gamma\sigma}}]}
  \ave{T[\crea{c_{-\vec{k}\alpha\dn}}(\tau)
    \crea{c_{\vec{k}'\beta\sigma}}]} \nonumber\\
  &= -\ave{T[\crea{c_{\vec{k}-\vec{q}\alpha\up}}(\tau)
    \crea{c_{-\vec{k}+\vec{q}\beta\dn}}]}
  \ave{T[\crea{c_{-\vec{k}\alpha\dn}}(\tau)
    \ani{c_{-\vec{k}\gamma\dn}}
  ]} \delta_{\sigma,\dn}\delta_{\vec{k}',-\vec{k}+\vec{q}} +
  \ave{T[\crea{c_{\vec{k}-\vec{q}\alpha\up}}(\tau)
    \ani{c_{\vec{k}-\vec{q}\gamma\up}}]}
  \ave{T[\crea{c_{-\vec{k}\alpha\dn}}(\tau)
    \crea{c_{\vec{k}\beta\up}}]} \delta_{\sigma,\up}
  \delta_{\vec{k}',\vec{k}}
\end{align}
\end{widetext}
Plugging this into Eq.~\eqref{eq.ex1}, we get from the first term
\begin{align}
  -\sum_{\vec{k}\alpha\beta\gamma}
  (-&G_{10}^{\beta\alpha}(-\tau,\vec{k}-\vec{q}))
  [\delta_{\mu}\Delta]^{\alpha\alpha}_{01} \times \nonumber \\
  \times &G_{11}^{\alpha\gamma}(\tau,\vec{k}) 
   (-[\partial_{\nu}\widetilde{H}|_{\vec{k}-\vec{q}/2}\gamma^{z}])^{\gamma\beta}_{11} 
  \nonumber\\
  = -\sum_{\vec{k}\alpha\beta\gamma}
  &G_{10}^{\beta\alpha}(-\tau,\vec{k}) 
  [\delta_{\mu}\Delta]^{\alpha\alpha}_{01}
  G_{11}^{\alpha\gamma}(\tau,\vec{k}+\vec{q})
  [\partial_\nu
    \widetilde{H}|_{\vec{k}+\vec{q}/2}\gamma^z]_{00}^{\gamma\beta}, 
\end{align}
where the transformation $\vec{k}\to\vec{k}+\vec{q}$ was used. Note that 
\begin{equation}
\partial_{\mu}\widetilde{H}|_{\vec{k}} = 
\begin{pmatrix}
    \frac{\partial H_{\up}(\vec{k}')}{\partial k_{\mu}'}\bigg|_{\vec{k}'=\vec{k}} & 0 \\
    0 & \frac{\partial H_{\dn}^*(\vec{k}')}{\partial k_{\mu}'}\bigg|_{\vec{k}'=-\vec{k}}
\end{pmatrix}.
\end{equation}
Similarly, the second part yields
\begin{align}
  &\sum_{\vec{k}\alpha\beta\gamma}
  -G_{00}^{\gamma\alpha}(-\tau,\vec{k}-\vec{q})
  [\delta_{\mu} \Delta]^{\alpha\alpha}_{01}
  G_{10}^{\alpha\beta}(\tau,\vec{k})
  [\partial_{\nu} \widetilde{H}|_{\vec{k}-\vec{q}/2}]_{00}^{\beta\gamma} 
  \nonumber
  \\
  &=
  -\sum_{\vec{k}\alpha\beta\gamma}
  G_{00}^{\gamma\alpha}(-\tau,\vec{k})
  [\delta_{\mu} \Delta]^{\alpha\alpha}_{01}
  G_{10}^{\alpha\beta}(\tau,\vec{k}+\vec{q})
  [\partial_{\nu} \widetilde{H}|_{\vec{k}+\vec{q}/2}]_{00}^{\beta\gamma}
\end{align}

Repeating this procedure for all terms involving one derivative of
$\Delta$ or $\Delta^*$, the total contribution is found to be
\begin{align}
  \Pi^{(1)}_{\mu\nu} &= -\sum_{\vec{k}} {\rm
    Tr}[G(-\tau,\vec{k})\delta_{\mu}\Delta 
    G(\tau,\vec{k}+\vec{q})
    \partial_{\nu}\widetilde{H}|_{\vec{k}+\vec{q}/2}\gamma^z] \nonumber
  \\
  &- \sum_{\vec{k}} {\rm Tr}[G(-\tau,\vec{k})
    \partial_{\mu} \widetilde{H}|_{\vec{k}+\vec{q}/2}\gamma^z
    G(\tau,\vec{k}+\vec{q})
    \delta_{\nu}\Delta] . \label{eq.contr1}
\end{align}

The next contributions to the paramagnetic current come from terms
which have a product of derivatives of $\Delta$ or $\Delta^*$ as a
prefactor, for example
\begin{equation}
  \sum_{\vec{k}\vec{k}'\alpha\beta} \frac{\delta
    \Delta_{\alpha}}{\delta A_{\mu}} \frac{\delta
    \Delta_{\beta}}{\delta A_{\nu}} \ave{
    T[ \crea{c_{\vec{k}-\vec{q}\alpha\up}} (\tau)
      \crea{c_{-\vec{k}\alpha\dn}} (\tau)
    \crea{c_{\vec{k}'+\vec{q}\beta\up}} \crea{c_{-\vec{k}'\beta\dn}} ]
    } .
\end{equation}

Like before, the correlator can be expressed as
\begin{align}
 &\ave{ T[ \crea{c_{\vec{k}-\vec{q}\alpha\up}} (\tau)
      \crea{c_{-\vec{k}\alpha\dn}} (\tau)
      \crea{c_{\vec{k}'+\vec{q}\beta\up}} \crea{c_{-\vec{k}'\beta\dn}}
  ] } \nonumber \\
  &= -\ave{T[ \crea{c_{\vec{k}-\vec{q}\alpha\up}}(\tau)
      \crea{c_{\vec{k}'+\vec{q}\beta\up}}]}
  \ave{
    T[ \crea{c_{-\vec{k}\alpha\dn}}(\tau)
      \crea{c_{-\vec{k}'\beta\dn}} 
    ]
  } \nonumber \\
  &+ \ave{T[ \crea{c_{\vec{k}+\vec{q}\alpha\dn}}(\tau)
      \crea{c_{\vec{k}'+\vec{q}\beta\up}} ]}
  \ave{T[ \crea{c_{\vec{k}-\vec{q}\alpha\up}}(\tau)
      \crea{c_{-\vec{k}'\beta\dn}}]} \nonumber \\
  &= \ave{T[\crea{c_{-\vec{k}\alpha\dn}}(\tau)
      \crea{c_{\vec{k}\beta\up}} 
  ]}
  \ave{
    T[\crea{c_{\vec{k}-\vec{q}\alpha\up}}(\tau)
      \crea{c_{\vec{q}-\vec{k}\beta\dn}}
    ]
  }\delta_{\vec{k}',\vec{k}-\vec{q}},
\end{align}
and the contribution to the paramagnetic term is
\begin{align}
  &\sum_{\vec{k}\alpha\beta}
  (-G_{10}^{\beta\alpha}(-\tau,\vec{k}-\vec{q}))
  [\delta_{\mu}\Delta]_{01}^{\alpha\alpha}
  G_{10}^{\alpha\beta}(\tau,\vec{k})
  [\delta_{\nu}\Delta]_{01}^{\beta\beta} \nonumber \\
  &=
  -\sum_{\vec{k}\alpha\beta}
  G_{10}^{\beta\alpha}(-\tau,\vec{k})
  [\delta_{\mu}\Delta]_{01}^{\alpha\alpha}
  G_{10}^{\alpha\beta}(\tau,\vec{k}+\vec{q})
  [\delta_{\nu}\Delta]_{01}^{\beta\beta} .
\end{align}
Repeating this for the other terms, the contribution to the
paramagnetic current is found to be
\begin{equation}
  \Pi^{(2)}_{\mu\nu} = -\sum_{\vec{k}} {\rm
    Tr}[G(-\tau,\vec{k})\delta_{\mu}\Delta G(\tau,\vec{k}+\vec{q})
  \delta_{\nu}\Delta].\label{eq.contr2}
\end{equation}
The last scalar term in the generalized paramagnetic current operator (Eq.~\eqref{eq.gen_para}) does not contribute, as it commutes with all operators. 

By combining equations~\eqref{eq.contr1} and~\eqref{eq.contr2}, we obtain
\begin{align}
  \Pi_{\mu\nu}(\vec{q},\tau) &= -\sum_{\vec{k}} {\rm Tr}\big[
    G(-\tau,\vec{k})(
    \partial_{\mu}\widetilde{H}|_{\vec{k}+\vec{q}/2}\gamma^z +
    \delta_{\mu}\Delta)  \nonumber \\
    &G(\tau,\vec{k}+\vec{q} )
    (\partial_{\nu}\widetilde{H}|_{\vec{k}+\vec{q}/2)}\gamma^z +
    \delta_{\nu}\Delta) 
    \big].
\end{align}

Fourier transforming $\Pi_{\mu\nu}$ to Matsubara space yields
\begin{align}
  \Pi_{\mu\nu}(\vec{q},i\omega_n) &= -\int_0^{\beta} {\rm d}\tau
  e^{i\omega_n\tau}\Pi_{\mu\nu}(\vec{q},\tau) \\
  &= \frac{1}{\beta}\sum_{\vec{k}}\sum_{\Omega_n} {\rm Tr}\big[
    G(i\Omega_n,\vec{k})(
    \partial_{\mu}\widetilde{H}|_{\vec{k}+\vec{q}/2}\gamma^z +
    \delta_{\mu}\Delta) \nonumber \\
    &G(i\Omega_n+i\omega_n,\vec{k}+\vec{q} )
    (\partial_{\nu}\widetilde{H}|_{\vec{k}+\vec{q}/2}\gamma^z +
    \delta_{\nu}\Delta) 
    \big] ,
\end{align}
where $\Omega_n=\pi (2n+1)/\beta$ is a fermionic Matsubara frequency and $\omega_n = 2\pi n/\beta$ is
a bosonic one. Computing the diamagnetic contribution to the current is
straightforward. The total current-current response function is given by

\begin{widetext}
  \begin{align}
    K_{\mu\nu}(\vec{q},i\omega_n) &= -\frac{1}{V}\frac{1}{\beta}
    \sum_{\vec{k}}\sum_{\Omega_m} {\rm Tr}\left[
      G(i\Omega_m,\vec{k})
      \partial_{\mu}\widetilde{H}|_{\vec{k}} 
      G(i\Omega_m,\vec{k} )
      \partial_{\nu}\widetilde{H}|_{\vec{k}}
      \right] \nonumber \\
    &+ \frac{1}{V}\frac{1}{\beta}\sum_{\vec{k}}\sum_{\Omega_n} {\rm
      Tr}\left[ 
      G(i\Omega_m,\vec{k})(
      \partial_{\mu}\widetilde{H}|_{\vec{k}+\vec{q}/2}\gamma^z +
      \delta_{\mu}\Delta) 
      G(i\Omega_m+i\omega_n,\vec{k}+\vec{q} )
      (\partial_{\nu}\widetilde{H}|_{\vec{k}+\vec{q}/2}\gamma^z +
      \delta_{\nu}\Delta) 
      \right] \nonumber \\
    &- \frac{1}{V_c}C\delta(\omega_n), \\
    C &= \frac{1}{U} \sum_{\alpha}
    \frac{\delta \Delta_{\alpha}}{\delta A_{\mu}} \frac{\delta
      \Delta_{\alpha}^*}{\delta A_{\nu}} + {\rm H.c.} 
  \end{align}
\end{widetext}

In mean-field theory, the BdG Hamiltonian can be diagonalized as
$H_{\rm BdG}=\sum_{a}E_a\ket{\psi_a}\bra{\psi_a}$, and the Green's function is
\begin{equation}
  G(i\Omega_n,\vec{k}) = \sum_{a}\frac{\ket{\psi_a}\bra{\psi_a}}
  {i\Omega_n-E_a(\vec{k})}. 
\end{equation}
The superfluid weight then becomes
\begin{widetext}
\begin{align}
  D_{s,\mu\nu} &= \lim_{\vec{q}\to 0} \lim_{\omega\to 0}
  K_{\mu\nu}(\vec{q},\omega)|_{\vec{A}=\vec{0}} \\
  &= \frac{1}{V}\sum_{\vec{k},a,b} \frac{n_F(E_b)-n_F(E_a)}{E_a-E_b} \big[
  \bra{\psi_a}\partial_{\mu}\widetilde{H}_{\vec{k}}\ket{\psi_b}
  \bra{\psi_b}\partial_{\nu} \widetilde{H}_{\vec{k}} \ket{\psi_a}
  \nonumber \\
  &-
  \bra{\psi_a}(\partial_{\mu}\widetilde{H}_{\vec{k}}\gamma^z
  +\delta_{\mu}\Delta)\ket{\psi_b}
  \bra{\psi_b}(\partial_{\nu}\widetilde{H}_{\vec{k}}\gamma^z
  +\delta_{\nu}\Delta)\ket{\psi_a}
  \big] - \frac{1}{V_c}C,
\end{align}
\end{widetext}
where $n_F(E)=1/(e^{\beta E}+1)$ is the Fermi-Dirac distribution and
the prefactor should be understood as $-\partial_E n_F(E)$ if
$E_a=E_b$. The functional derivatives of the order parameters can be computed with knowledge of only the ground state at $\vec{A}=\vec{0}$, for example by using the Hessian method presented in the main text (see Eq.~\ref{eq.lin_sys_del}).

These equations are valid in general as long as $\delta\mu/\delta A_{\mu}\big|_{\vec{A}=\vec{0}} = 0$. If the derivative of the chemical potential is not zero, the above will be equivalent to $D_s=(1/V){\rm d}^2\Omega/{\rm d}q_i{\rm d}q_j\big|_{\mu,\vec{q}=\vec{0}}$, where $\mu$ is kept constant when taking the derivative. This may not be equal to $(1/V){\rm d}^2F/{\rm d}q_i{\rm d}q_j\big|_{N,\vec{q}=\vec{0}}$, where the particle number is kept constant.

\begin{figure}
  \centering
  \includegraphics[width=\columnwidth]{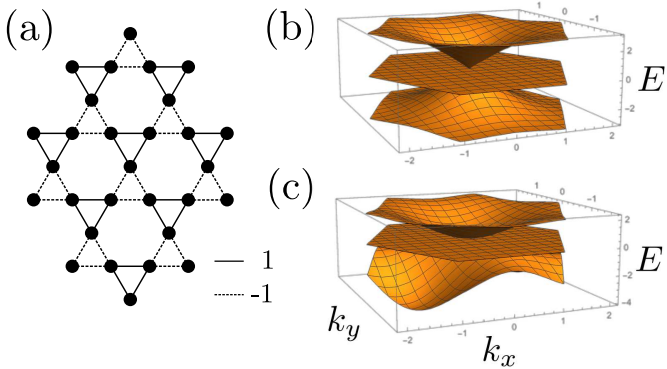}
  \caption{(a) Schematic representation of the kagome model with a
    linear band touching. The corresponding band structure is shown in
  (b). (c) Band structure of the corresponding model with a quadratic
    band touching.}\label{fig.kago}
\end{figure}

\section{Equivalence of $D_s$ obtained from the thermodynamic
  potential and linear response theory} \label{app.equivalence}

In this Appendix, we will show that the definition $[D_s]_{\mu\nu} = (1/V){\rm
  d}^2 F/{{\rm d}q_{\mu}{\rm d}q_{\nu}}\big|_{\vec{q}=\vec{0}}$ is equivalent
to the result obtained from linear response theory, $[D_s]_{\mu\nu} =
\lim_{\vec{q}\to 0}\lim_{\omega\to 0} K_{\mu\nu}(\vec{q},\omega)
\big|_{\vec{A}=\vec{0}}$, where $K_{\mu\nu}$ is the current-current
response function, $\ave{j_{\mu}(\vec{q},\omega)} =
-K_{\mu\nu}(\vec{q},\omega)A_{\nu}(\vec{q},\omega)$. 

When we define $[D_s]_{\mu\nu} = (1/V){\rm d^2}F/{{\rm d}q_{\mu}{\rm
    d}q_{\nu}}\big|_{\vec{q}=\vec{0}}$, the vector $\vec{q}$ is introduced
in the phase of the order parameters $\Delta_{i\alpha}\to
\Delta_{i\alpha}e^{2i\vec{q}\cdot\vec{r}_{i\alpha}}$. This phase can be
moved to the kinetic Hamiltonian with a unitary transformation
$c_{i\alpha\sigma}\to c_{i\alpha\sigma}e^{-i\vec{q}\cdot
  \vec{r}_{i\alpha}}$. The vector $\vec{q}$ is thus equivalent to a
constant vector potential $\vec{A}$ introduced via a Peierls substitution.

The grand canonical potential is defined as $\Omega(\vec{A}) =
-\beta^{-1} \ln Z(\vec{A})$, $Z(\vec{A}) = {\rm Tr}\left[ e^{-\beta
    H(\vec{A})}\right]$. The term $\mu N$ is included in the
Hamiltonian, see Eq.~\eqref{eq.ham}. The functional derivative of
$\Omega$ 
is 
\begin{align}
  \frac{1}{V}\frac{\delta^2\Omega}{\delta A_{\mu}\delta A_{\nu}} &=
  \frac{1}{VZ} \frac{\delta}{\delta A_{\mu}}{\rm Tr}\left[ \frac{\delta
      H}{\delta A_{\nu}} e^{-\beta H(\vec{A})} \right] \nonumber
  \\
  &=
  \frac{\delta}{\delta A_{\mu}} \sum_{\lambda} K_{\nu\lambda}A_{\lambda} = K_{\nu\mu}. 
\end{align}
Thus
\begin{align}
  [D_{s}]_{\mu\nu} &= \frac{1}{V}\frac{{\rm d}^2F}{{\rm d}A_{\mu}{\rm d}A_{\nu}}
  \bigg|_{\vec{A}=\vec{0},N} = \frac{1}{V}\frac{{\rm d}^2\Omega}{{\rm
      d}A_{\mu}{\rm d}A_{\nu}}\bigg|_{\vec{A}=\vec{0},\mu} \nonumber \\
  &=\lim_{\vec{q}\to\vec{0}} \lim_{\omega \to 0} K_{\mu\nu}(\vec{q},\omega),
\end{align}
assuming that the chemical potential has a vanishing derivative at $\vec{A}=0$. When taking the total
derivative of $F$, the total particle number is kept
constant, whereas for $\Omega$, the chemical
potential is kept constant.

\section{Flat band models with a tuned band touching}~\label{app.ltq}

In Sec.~\ref{sec.ltq}, we presented results for flat band models with a
tuned band touching. We used the method developed in
Ref.~[\onlinecite{Graf2021}] to construct models where the flat band energy
and eigenstates remain unchanged while the band touchings with
the dispersive bands are tuned from linear to quadratic. For the kagome
geometry, the model with a linear band touching is shown in
Fig.~\ref{fig.kago}. The Fourier transformed kinetic Hamiltonian is
\begin{equation}
  H_{\vec{k},\rm lin, kago} = -2i\begin{pmatrix}
  0 & \sin(k_1/2) & \sin(k_2/2) \\
  -\sin(k_1/2) & 0 & -\sin(k_3/2) \\
  -\sin(k_2/2) & \sin(k_3/2) & 0
  \end{pmatrix},
\end{equation}
where $k_1=k_x$, $k_2=k_x/2+\sqrt{3}k_y/2$ and $k_3 =
k_x/2-\sqrt{3}k_y/2$. The length of a unit cell lattice vector is taken equal to
$1$. This model has a flat band at $E=0$. The corresponding quadratic
model is constructed so that the flat band is at the same energy and
has the same Bloch functions. The obtained kinetic Hamiltonian is
\begin{widetext}
\begin{equation}
  H_{\vec{k},\rm quad,kago} = C\begin{pmatrix}
    \sin^2(k_1/2)+\sin^2(k_2/2) & -\sin(k_2/2)\sin(k_3/2) &
    \sin(k_1/2)\sin(k_3/2) \\
    -\sin(k_2/2)\sin(k_3/2) & \sin^2(k_3/2)- 2\sin^2(k_1/2) &
    -2\sin(k_1/2)\sin(k_2/2) \\
    \sin(k_1/2)\sin(k_3/2) & -2\sin(k_1/2)\sin(k_2/2) &
    \sin^2(k_3/2) - 2\sin^2(k_2/2)
  \end{pmatrix}.
\end{equation}
\end{widetext}
The constant $C$ is chosen so that the total width of the band
structure is the same as in the linear model. The obtained band
structure is shown in Fig.~\ref{fig.kago}c. The total Hamiltonian
with a continuously tuned band touching is $H_{\vec{k},\rm kago} =
[(1-\lambda)H_{\vec{k},\rm lin,kago} + \lambda H_{\vec{k},\rm
    quad,kago}]/C_2(\lambda)$, where $C_2$ is chosen so that the total
width of the band structure is independent of $\lambda$. Since both
the linear and quadratic model have a flat band at the same energy
with the same eigenfunctions, the flat band eigenstates are identical
for all $\lambda$.

For the Lieb geometry, we choose the same Lieb lattice as our linear
model. In order to be able to open a band gap, we introduce the 
staggered hopping amplitudes used in the main text. The kinetic Hamiltonian is
\begin{widetext}
\begin{equation}
  H_{\vec{k},\rm lin,Lieb} = -2\begin{pmatrix}
  0 & \cos(k_x/2) +i\delta\sin(k_x/2) & \cos(k_y/2)+i\delta\sin(k_y/2)
  \\
  \cos(k_x/2) -i\delta\sin(k_x/2) & 0 & 0 \\
  \cos(k_y/2) -i\delta\sin(k_y/2) & 0 & 0 
  \end{pmatrix}.
\end{equation}
\end{widetext}
The kinetic Hamiltonian for the corresponding quadratic model is
\begin{widetext}
\begin{equation}
  H_{\vec{k},\rm quad,Lieb} = -\frac{1}{\sqrt{2}} \begin{pmatrix}
    -2(1+\delta^2) - (1-\delta^2)(\cos(k_x)+\cos(k_y)) & 0 & 0 \\
    0 & 1+\delta^2 + (1-\delta^2)\cos(k_x) & \Lambda(k_x,k_y,\delta) \\
    0 & \Lambda^{*}(k_x,k_y,\delta) & 1+\delta^2+(1-\delta^2)\cos(k_y)
  \end{pmatrix}, \label{eq:quadratic_lieb}
\end{equation}
\end{widetext}
where $\Lambda(k_x,k_y,\delta) = 2(\cos k_x/2-i\delta\sin k_x/2)(\cos k_y/2 +i\delta\sin k_y/2)$. 
One sublattice is disconnected from the others in this model. In this case, fixing the phase of one order parameter is not sufficient to make the Hessian matrix invertible, and we need to fix the overall phase in both the disconnected sublattice and the remaining two-band model. The
total interpolating Hamiltonian is obtained the same way as for the kagome
lattice. In this case, at $\delta=0$, the band touching is tuned
continuously from a linear to a quadratic one. For nonzero $\delta$, a gap is opened. In this case, tuning $\lambda$ modifies the dispersive bands are modified without affecting the geometry of the flat band.

\section{S-Matrix construction} \label{app.smatrix}

The $S$-matrix bipartite Hamiltonians\cite{Calugaru2021} offer a route to understanding the mean-field gap in flat band systems, even those with band touching points.  Denote the two sublattices $L, \tL$ with $N_L > N_\tL$, where $N_L, N_\tL$ are the number of orbitals per unit cell of each sublattice~\cite{Calugaru2021}.  The kinetic energy Hamiltonian reads
\begin{align}
H_\kk = \begin{bmatrix}
0 & S_\kk^\dagger \\
S_\kk & 0
\end{bmatrix}.
\label{}
\end{align} Here $S_\kk^\dagger$ is an $N_\tL \times N_L$ dimensional matrix, and so has a $N_L - N_\tL$ dimensional null space that forms the flat bands.  This Hamiltonian obeys a chiral symmetry
\begin{align}
S = \begin{bmatrix}
I_{\tL \times \tL} & 0 \\
0 & -I_{L \times L}
\end{bmatrix}, ~~\{S, H_\kk\} = 0,
\label{}
\end{align} and the dispersive and flat wavefunctions read
\begin{align}
\Psi^\text{disp}_{\kk,m,\pm} = \dfrac{1}{\sqrt{2}} \begin{bmatrix}
\phi_{\kk,m} \\ \pm \psi_{\kk,m}
\end{bmatrix},~~  \Psi^\text{flat}_{\kk,n} = \begin{bmatrix}
  0 \\ \psi_{\kk,n}
  \end{bmatrix}.
\label{}
\end{align}  Here $\phi_{\kk,m}$ and $\psi_{\kk,m}$ are normalized column vectors whose components correspond the orbitals in the $\tilde L$ and $L$ sublattices, respectively. The vector $\phi_{\kk,m}$ has length $N_\tL$, and $\psi_{\kk,m}$ has length $N_L$.  The dispersive states have energy $\pm \epsilon_{\kk,m}$, where $\epsilon_{\kk,m}$ are the singular values of $S_\kk$.   Due to chiral symmetry, the $\phi$ and $\psi$ sublattice vectors obey their own orthonormality relations and one may define the sublattice projectors as 
\begin{align}
P^\tL_{m}(\kk) = \phi_{\kk,m} \phi_{\kk,m}^\dagger,~P^L_{m}(\kk) = \psi_{\kk,m} \psi_{\kk,m}^\dagger,
\label{}
\end{align}  where $P^{\tL}_m(\kk)$ is a $\tL \times \tL$ dimensional matrix, running over the orbitals $\alpha$ in the smaller sublattice $\tL$, and $P^{L}_m(\kk)$ is an $L \times L$ dimensional matrix, running over the orbitals $\alpha$ in the larger sublattice $L$.  We allow the index $m$ to run over {\it both} the $N_\tL$ positive energy dispersive bands and the $N_L - N_\tL$ flat bands.  Because there is no weight of the wavefunction in the smaller sublattice $\tL$ in the flat bands,  $P^\tL_m(\kk) = 0$ for $m$ in the flat bands.  The sublattice projectors satisfy
\begin{align}
\text{Tr}[P^\tL_m(\kk)] &= \begin{cases}
0 & \text{~if~} m \in \text{flat bands} \\
1 & \text{~if~} m \in \text{dispersive bands}
\end{cases},\\
\text{Tr}[P^L_m(\kk)] &= 1.
\label{eq:sublattice_projectors}
\end{align}  These projectors are Hermitian and square to themselves, as expected.

\subsection{Linear and quadratic band touchings}
The $S$-matrix Hamiltonian lends itself naturally to construct models with linear band touchings at high-symmetry momenta \cite{Calugaru2021}, and the quadratic band touchings can be derived in a simple manner.  Consider the new Hamiltonian 
\begin{align}
H_\text{quad} = \begin{bmatrix}
-S_\kk^\dagger S_\kk & 0 \\
0 & S_\kk S^\dagger_\kk
\end{bmatrix},
\label{eq:ham_quad}
\end{align} where $S_\kk S^\dagger_\kk$ is the line graph derived from $L, \tL$ \cite{Calugaru2021}.  If $H_\kk$ has a linear band touching point, then $H_\text{quad}$ has quadratic band touchings. While the flat band wavefunctions of $H_\text{quad}$ are the same as $H_\kk$, the dispersive wavefunctions change.  This quadratic construction is precisely the construction employed in the Lieb lattice quadratic band touching point discussed in Appendix~\ref{app.ltq}.  While the quadratic band touching point breaks chiral symmetry, the wavefunctions are still expressed in terms of the sublattice vectors $\phi, \psi$, allowing for a precise treatment of the self-consistent mean field gap equations.

\section{S-Matrix mean field theory} \label{app.smatrixmft}

Adding the Hubbard interaction to the $S$-matrix and performing a mean-field analysis yields the BdG Hamiltonian
\begin{align}
H_\text{MF} &= \sum_{\kk,\sigma,\alpha\beta} [H_\kk]_{\alpha\beta} c^\dagger_{\kk,\alpha,\sigma} c_{\kk,\beta,\sigma} \nonumber \\
&+ \sum_{\kk,\alpha} \Delta_{\alpha} c^\dagger_{\kk,\alpha,\uparrow} c^\dagger_{-\kk,\alpha,\downarrow} + H.c.,
\label{}
\end{align} where
\begin{align}
\Delta_{\RR {\alpha}} = U \ave{ c_{\RR,\alpha,\downarrow} c_{\RR,\alpha,\uparrow}} = \Delta_\alpha,
\label{}
\end{align} with the Hubbard interaction parameter $U < 0$, translation invariance in $\Delta_{\RR \alpha} = \Delta_\alpha$, and $U(1)_z$-spin conservation and time reversal symmetry.  Further, we assume uniform pairing within each sublattice: $
\Delta_\alpha = \Delta_L$ or $\Delta_\tL$ depending on the sublattice $\alpha$ belongs to.  Such a condition may be enforced by symmetries that relate each orbital within each sublattice \cite{upcoming}.

\subsection{Linear band touching (with chiral symmetry)}
Using the non-redundant BdG basis the Hamiltonian is expressed as
\begin{align}
H^\text{BdG}_\kk = \begin{bmatrix}
0 & S_\kk^\dagger & \Delta_{\tilde L} I_{\tL \times \tL} & 0 \\
S_\kk & 0 & 0 & \Delta_{L} I_{L \times L} \\
\Delta_{\tilde L} I_{\tL \times \tL} & 0 & 0 & -S_\kk^\dagger \\
0 & \Delta_{L} I_{L \times L} & -S_\kk & 0
\end{bmatrix}.
\label{}
\end{align} The BdG Hamiltonian possesses a chiral symmetry arising from the product of TRS and particle-hole.  There is another chiral symmetry inherited from the bipartite lattice.  The product of these two symmetries yields a unitary symmetry.

This Hamiltonian can be solved exactly and the positive energy eigenvalues read
\begin{align}
E_{\kk,m}^{1,2} = \dfrac{1}{2} \left[\pm(\Delta_\tL - \Delta_L) + \sqrt{(\Delta_\tL + \Delta_L)^2 + 4\epsilon_{\kk,m}^2} \right],~~ \label{eq:energy_chiral}
\end{align}  In the situation where $m$ is a flat band,
\begin{align}
E_{\kk,m} = \Delta_L.
\label{}
\end{align}

If the kinetic Hamiltonian possesses a band touching point arising from symmetry, the degeneracy between the flat bands and dispersive bands will be made manifest in the BdG spectrum.  Assume that at high symmetry momentum $\KK$, the flat bands and band touching points transform under representation $X \oplus Y$, where $X$ is the representation induced by orbitals in the $L$ sublattice, and $Y$ the representation induced by orbitals in the $\tL$ sublattice.  The dimensions obey $\text{dim}(X) - \text{dim}(Y) = N_L - N_\tL, \text{dim}(Y) > 0$.  When pairing is added, those bands transforming under irrep $X$ gain energy $\pm \Delta_L$, and there are $\text{dim}(Y)$ bands in addition to the flat bands that are degenerate.  These new band-touching points are quadratic.

\subsection{Quadratic band touching (no chiral symmetry)}
The quadratic band touching Hamiltonian Eq.~\ref{eq:ham_quad} no longer possesses chiral symmetry, but it does factor into sublattices $\tL, L$.  This is true even when pairing is added:

\begin{align}
H^\text{BdG}_\text{quad} = \begin{bmatrix}
- S_{\kk}^\dagger S_{\kk} & 0 & \Delta_\tL I_{\tL \times \tL} & 0 \\
0 & S_{\kk} S_{\kk}^\dagger  & 0 & \Delta_L I_{L \times L} \\
\Delta_\tL I_{\tL \times \tL} & 0 & S_{\kk}^\dagger S_{\kk} & 0 \\
0 & \Delta_L I_{L \times L} & 0 & -S_{\kk} S_{\kk}^\dagger
\end{bmatrix}.
\label{}
\end{align} 
 Thus, each sublattice may be treated separately.
The positive energy eigenvalues are
\begin{align}
E_{\kk,m,\tL} = \sqrt{\epsilon_{\kk,m}^4 + \Delta_\tL^2}, ~~ E_{\kk,m,L} = \sqrt{ \epsilon_{\kk,m}^4 + \Delta_L^2},
\label{}
\end{align} 
Though chiral symmetry no longer holds, this quadratic Hamiltonian still possesses the band touching point at energy $\pm \Delta_L$.

\section{Gap equation} \label{app.gap}

\subsection{Chiral symmetric Hamiltonian}
For the chiral symmetric Hamiltonian $H_\kk$, the gap equation at zero temperature \cite{Peotta2015} reads 
\begin{align}
\Delta_\alpha &= \frac{|U|}{N} \sum_{\kk,m} \dfrac{\Delta_L + \Delta_{\tL}}{2 \sqrt{(\Delta_L + \Delta_{\tL})^2 + 4\epsilon_{\kk,m}^2}} \nonumber \\
&~~~~~~~\times [P^{(L)}_{m}(\kk) \oplus P^{(\tilde L)}_{m}(\kk)]_{\alpha\alpha}.
\label{eq:gap}
\end{align}

Employing the trace relations Eq.~\ref{eq:sublattice_projectors} removes the projectors and all wavefunction dependence, yielding gap equations
\begin{align}
N_L \Delta_L &= \frac{|U|N_\tL}{2} f(\Delta) + \frac{|U|(N_L - N_\tL)}{2}\\
N_\tL \Delta_\tL &= \frac{|U|N_\tL}{2} f(\Delta).
\label{eq:linear_gaps}
\end{align}   where we have defined
\begin{align}
f(\Delta) &= \dfrac{1}{N N_\tL} \sum_{\kk, m \in \text{disp}}  \dfrac{\Delta}{ \sqrt{\Delta^2 + \epsilon_{\kk,m}^2}} \\
\Delta &= \dfrac{1}{2}(\Delta_L + \Delta_\tL).
\end{align}

This leads to the {\it{universal}} relation
\begin{align}
N_L \Delta_L - N_\tL \Delta_\tL = \frac{|U|(N_L - N_\tL)}{2},
\label{}
\end{align} where we recognize the RHS of this weighted difference equation as the strength of pairing arising from the flat bands.  This equation is universal is it only requires the bipartite nature of the underlying model, and does not depend on the dispersion or wavefunctions, nor the presence or absence of band touching points.  We have verified the weighted difference relation numerically, and the relation has been seen to hold in the Lieb lattice~\cite{Julku2016}.

Further bounds on $\Delta_L, \Delta_\tL$ can be proven by noting that $f(\Delta)$ is monotonically increasing in $\Delta$ and ranges from $0$ to $1$.  The gap equation for the average pairing gap $\Delta$ reads
\begin{align}
\frac{\Delta}{U} &= \dfrac{1}{4}(1+r) f(\Delta) + \dfrac{1}{4}(1-r), ~r = \dfrac{N_\tL}{N_L}
\label{}
\end{align}  which only depends on the average form of the dispersive bands $f(\Delta)$ and the ratio of the sublattice orbital numbers $r$.  This equation always has a solution, as the right hand side is positive and bounded.  As $0 < f(\Delta) < 1$, the average pairing obeys
\begin{align}
    \frac{1}{4}(1-r) < \frac{\Delta}{|U|} < \frac{1}{2}.
\end{align}
The pairing on the $L$ sublattice is always larger than the pairing on the $\tL$ sublattice, $\Delta_L > \Delta_\tL$:
\begin{align}
\frac{\Delta_L}{|U|} - \frac{\Delta_\tL}{|U|} &= \frac{1}{2}(1-r)(1-f(\Delta)) > 0. 
\label{}
\end{align} 
The larger pairing $\Delta_L$ is maximized when $r \rightarrow 0$, or if the ratio of flat bands to dispersive bands is made as large as possible.  Because there is a solution $\Delta > 0$, it follows from the self-consistent equations that $\Delta_L, \Delta_\tL > 0$, i.e. there is pairing on both sublattices:
\begin{align}
    \Delta_\tL > 0,~\Delta_L = \frac{|U|}{2} (1-r) + r \Delta_\tL.
\end{align}

\subsection{Quadratic band touching}
The gap equations for the quadratic Hamiltonian decouple into $\tL, L$ sectors.  Defining
\begin{align}
f^\text{quad}(\Delta) &= \dfrac{1}{N N_\tL} \sum_{m \in \text{disp}} \sum_\kk \dfrac{\Delta}{\sqrt{\Delta^2 + \epsilon_{\kk,m}^4}}, \\
\end{align}  the self-consistent equations read
\begin{align}
\Delta_\tL &= \frac{|U|}{2} f^\text{quad}(\Delta_\tL) \\
\Delta_L &= \frac{|U|}{2} r f^\text{quad}(\Delta_L) + \dfrac{|U|}{2}(1-r).
\label{eq:selfconquad}
\end{align}  As in the linear case, this system always has at least one solution:  $\Delta_\tL = 0$ satisfies the first equation and the second always has a solution as $f^\text{quad}(\Delta_L)$ is bounded.

The universal relation for the chiral Hamiltonians no longer holds (as the quadratic band touching model does not obey the chiral symmetry): instead the weighted difference reads
\begin{align}
    N_L \Delta_L - N_\tL \Delta_\tL &= \frac{|U| N_\tL}{2} [f^\text{quad}(\Delta_L) - f^\text{quad}(\Delta_\tL)] \nonumber \\
    &+ \frac{|U|(N_L - N_\tL)}{2}.
\end{align} If $\Delta_L > \Delta_\tL$, then regardless of the form of $f^\text{quad}$, the weighted pairing difference $N_L \Delta_L - N_\tL \Delta_\tL$ increases from the linear model to the quadratic model.  In the linear model it is clear that $\Delta_L > \Delta_\tL$, and if $\Delta_\tL = 0$ in the quadratic model, the inequality is also obvious.

Unfortunately, one cannot make the general claim that $\Delta_L > \Delta_\tL$.  Though we expect $\Delta_L > \Delta_\tL$ as the flat bands contribute to the superconductivity in $\Delta_L$ but not $\Delta_\tL$, we can only prove the slightly weaker statement: if there is a self-consistent solution $\Delta_\tL$, there is also a self-consistent solution $\Delta_L$ where $\Delta_L > \Delta_\tL$.  To prove this, define the functions 
\begin{align}
    u(\Delta) &= \frac{|U|}{2} f^\text{quad}(\Delta), \\
    v(\Delta) &= \frac{|U|}{2} r f^\text{quad}(\Delta) + \frac{|U|}{2}(1-r),
\end{align} 
and as such $0 < u(\Delta) < v(\Delta) < \frac{|U|}{2}$. Assume the fixed point $u(\Delta_\tL) = \Delta_\tL$.  Because $u(\Delta) < v(\Delta)$, we have
\bea
\Delta_\tL - v(\Delta_\tL) < \Delta_\tL - u(\Delta_\tL) = 0 \ .
\eea
But note that the function $\Delta - v(\Delta)$ also attains positive value by setting $\Delta = |U|/2+\eps, \ \eps>0$ and using $v(\Delta)<|U|/2$. By the intermediate value theorem, there exists $\Delta_L \in (\Delta_\tL, |U|/2)$ such that $\Delta_L - v(\Delta_L) = 0$. Hence we have demonstrated a solution exists to \Eq{eq:selfconquad} where $\Delta_L > \Delta_\tL$.  This establishes that there exists a self-consistent solution of \Eq{eq:selfconquad} where $\Delta_L > \Delta_\tL$ and thus $N_L \Delta_L - N_\tL \Delta_\tL$ increases in the quadratic band touching case relative to the linear band touching case, though this is due to the nature of the dispersive band wavefunctions and not the dispersion.

We emphasize that the universal relations between $\Delta_L, \Delta_\tL$ we have derived arise due to the geometry of the bipartite wavefunctions, and not due the dispersion.  This is a striking result of the bipartite $S$-matrix construction: various inequalities regarding the strength of the pairing gap can be made without recourse to the details of the model.  The details, however, do affect the physics: tuning the band touching point to be quadratic should enhance the gap $\Delta_L$, as the quadratic band structure has greater density of states at low energy, increasing $f(\Delta)$. 

\subsection{Connection to Lieb's theorem and the Uniform Pairing Models}
One can connect our mean field results to the models studied by Refs.~\cite{Lieb1989,Mielke1992, Mielke1993, Tovmasyan2016}.  In his seminal paper, Lieb proved that the ground state of a bipartite lattice with on-site attractive interactions, assuming appropriate symmetries, is unique.  If the flat bands are gapped from the dispersive bands, one can project away the dispersive bands and further argue that the ground state takes the form of the BCS wavefunction \cite{Julku2016} (this is not necessarily true if there are band touching points).  Because the dispersive bands have been projected away, there is no weight of the flat bands in the smaller sublattice, so $\Delta_\tL = 0$.  Our mean field results yield a particularly simple result in this projected limit: if the dispersive bands are sufficiently gapped from the flat bands, $f(\Delta) \rightarrow 0$, and the pairings in the sublattices read 
\begin{align}
\Delta_\tL = 0, ~~\Delta_L = \frac{|U|(N_L - N_\tL)}{2N_L}.
\end{align}  The strength of the pairing $\Delta_L$ is universal and does not depend on the form of the wavefunctions.

\bibliography{sources,sources2,bib_tbg,biblio_Nat_Phys_Rev_Sebastiano}

\end{document}